\documentclass[11pt]{article}

\usepackage[preprint]{acl}

\usepackage{times}
\usepackage{latexsym}

\usepackage[T1]{fontenc}

\usepackage[utf8]{inputenc}

\usepackage{microtype}

\usepackage{inconsolata}

\usepackage{graphicx}

\usepackage{multirow}
\usepackage{makecell}
\usepackage{booktabs}
\usepackage{tikz}

\usepackage{amsmath}
\usepackage{amssymb}
\usepackage{algorithm}
\usepackage{algorithmic}
\usepackage{bm}
\usepackage{bm}
\usepackage{bbold}

\usepackage{color, colortbl}  
\usepackage{graphicx}
\usepackage{enumitem}  
\usepackage{caption}

\usepackage{amsmath}
\usepackage{amssymb}
\usepackage{algorithm}
\usepackage{algorithmic}
\usepackage{bm}
\usepackage{bbold}

\usepackage{tcolorbox}
\tcbuselibrary{breakable}

\usepackage{natbib}

\usepackage{wasysym}
\usepackage{pifont} 
\definecolor{kellygreen}{rgb}{0.3, 0.73, 0.09}
\definecolor{alizarin}{rgb}{0.82, 0.1, 0.26}
\definecolor{royalblue}{rgb}{0.25,0.41,1}
\newcommand{\cmark}{{\color{kellygreen} \ding{51}}}
\newcommand{\xmark}{{\color{alizarin} \ding{55}}}

\newcommand{\best}[1]{{\textbf{#1}}}
\newcommand{\second}[1]{{\underline{#1}}}

\usepackage{xspace}
\newcommand\dataset{WebTestBench\xspace}
\newcommand\agent{WebTester\xspace}

\title{WebTestBench: Evaluating Computer-Use Agents towards \\End-to-End Automated Web Testing}

\author{
Fanheng Kong\textsuperscript{1,2\thanks{Work done during an internship at Kuaishou Technology.}}\; Jingyuan Zhang\textsuperscript{2}\; Yang Yue\textsuperscript{2}\; Chenxi Sun\textsuperscript{2}\; Yang Tian\textsuperscript{2} \\
\textbf{Shi Feng}\textsuperscript{1\thanks{Corresponding Author.}}\; \textbf{Xiaocui Yang}\textsuperscript{1}\; \textbf{Daling Wang}\textsuperscript{1}\; \textbf{Yu Tian}\textsuperscript{2}\; \textbf{Jun Du}\textsuperscript{2} \\
\textbf{Wenchong Zeng}\textsuperscript{2}\; \textbf{Han Li}\textsuperscript{2}\; \textbf{Kun Gai}\textsuperscript{2} \\
\vspace{-0.8em} \\
\textsuperscript{1}Northeastern University \quad
\textsuperscript{2}Kuaishou Technology \\
}

\begin{document}
\maketitle

\begin{abstract}

The emergence of Large Language Models (LLMs) has catalyzed a paradigm shift in programming, giving rise to ``vibe coding'', where users can build complete projects and even control computers using natural language instructions. This paradigm has driven automated webpage development, but it introduces a new requirement about how to automatically verify whether the web functionalities are reliably implemented. Existing works struggle to adapt, relying on static visual similarity or predefined checklists that constrain their utility in open-ended environments. Furthermore, they overlook a vital aspect of software quality, namely latent logical constraints. To address these gaps, we introduce \dataset, a benchmark for evaluating end-to-end automated web testing. \dataset encompasses comprehensive dimensions across diverse web application categories. We decompose the testing process into two cascaded sub-tasks, checklist generation and defect detection, and propose \agent, a baseline framework for this task. Evaluating popular LLMs with \agent reveals severe challenges, including insufficient test completeness, detection bottlenecks, and long-horizon interaction unreliability. These findings expose a substantial gap between current computer-use agent capabilities and industrial-grade deployment demands. We hope that \dataset provides valuable insights and guidance for advancing end-to-end automated web testing.
Our dataset and code are available at \url{https://github.com/friedrichor/WebTestBench}.

\end{abstract}

\section{Introduction}
\label{sec:introduction}

Web applications, serving as foundational platforms supporting a wide range of daily activities, provide rich services and content, becoming an indispensable pillar of modern society~\citep{commoncrawl2026}. As user demand grows, web pages proliferate rapidly, with increasing functionality and complexity. However, developing a web application is complicated and time-consuming, requiring developers to master diverse technical skills (\textit{e.g.}, CSS, JavaScript), and adapt to evolving frameworks (\textit{e.g.}, React, Vue)~\citep{borges2024skills}. Moreover, a complete development lifecycle is collaborative, spanning specialized roles such as coding, testing, and documentation~\citep{liang2022understanding}.

\begin{table*}[t]
\centering
\resizebox{\textwidth}{!}{
\begin{tabular}{lccccccccl}
\toprule
\textbf{Benchmark} & \textbf{\#Sample} & \textbf{Task} & \textbf{Env} & \textbf{Checklist} & \textbf{Inte.} & \textbf{L.H.} & \textbf{Func.} & \textbf{Cons.} & \textbf{Test Method}  \\
\midrule
\rowcolor{gray!10}\multicolumn{10}{l}{\textit{\textbf{Application Quality Evaluation Methods Originate from Web Generation Tasks}}}\\
Web2Code~\citep{yun2024web2code} & 60K & Generation & Web & \xmark & \xmark & \xmark & \xmark & \xmark & Visual \\
Interaction2Code~\citep{xiao2024interaction2code}  & 127 & Generation & Web & \xmark & \cmark & \xmark & \cmark & \xmark & Visual+Rule \\
Design2Code~\citep{si2025design2code} & 484 & Generation & Web & \xmark & \xmark & \xmark & \xmark & \xmark & Visual \\
DesignBench~\citep{xiao2025designbench} & 900 & Generation & Web & \xmark & \xmark & \xmark & \xmark & \xmark & Visual+Rule+MLLM \\ 
WebGen-Bench~\citep{lu2025webgenbench}  & 101 & Generation & Web & \xmark & \cmark & \cmark & \cmark & \xmark & Agent \\
ArtifactsBench~\citep{zhang2025artifactsbench}  & 1,825 & Generation & Web & \xmark & \cmark & \cmark & \cmark & \xmark & Visual+Script+MLLM \\
WebCoderBench~\citep{liu2026webcoderbench} & 1,572 & Generation & Web & \xmark & \cmark & \xmark & \cmark & \xmark & Rule+LLM \\
\midrule
\rowcolor{gray!10}\multicolumn{10}{l}{\textit{\textbf{Application Testing Benchmarks}}}\\
GTArena~\citep{zhao2024gtarena} & 10,844 & Testing & Mobile & \cmark & \cmark & \xmark & \cmark & \xmark & Agent \\
GUITestBench~\citep{gao2026guitester} & 143 & Testing & Mobile & \xmark & \cmark & \cmark & \cmark & \xmark & Agent \\
PlanATA~\citep{chevrot2025planata} & 113 & Testing & Web & \xmark & \cmark & \cmark & \cmark & \xmark & Agent \\
WebTestPilot~\citep{teoh2026webtestpilot} & 100 & Testing & Web & \xmark & \cmark & \cmark & \cmark & \xmark & Agent \\ 
\midrule
\dataset & 100 & Testing & Web & \cmark & \cmark & \cmark & \cmark & \cmark & Agent \\
\bottomrule
\end{tabular}
}
\caption{Comparison with representative related benchmarks across several aspects: the number of samples (\textbf{\#Sample}); the primary task (\textbf{Task}); the app environment (\textbf{Env}); whether the evaluation supports end-to-end testing without manually predefined test items (\textbf{Checklist}); whether testing requires dynamic interaction with the app (\textbf{Inte.}); whether it involves long‑horizon interactions (\textbf{L.H.}); whether functional correctness is evaluated (\textbf{Func.}); whether latent logical constraints are evaluated (\textbf{Cons.}); and the approach used to execute tests (\textbf{Test Method}).
}
\label{table:benchmark_comparison}
\vspace{-1em}
\end{table*}

The rise of LLMs has provided support for efficient coding~\citep{fried2023incoder, li2023starcoder, jiang2024survey}. Agentic coding products, exemplified by Claude Code~\citep{anthropic2026claudecode}, Gemini~\citep{google2026gemini}, and Codex~\citep{openai2026codex}, are rapidly gaining traction, enabling non-expert users to build complete web applications from scratch via natural language instructions, effectively compressing the traditional multi-person development team into a single human-AI collaborative unit. However, these products rarely generate flawless outcomes in a single attempt, and the synthesized applications frequently exhibit omissions or defects in both appearance and functionality~\citep{lu2025webgenbench, wan2025tddev}. For creators lacking software engineering expertise, verifying the quality and reliability of these generated applications remains a barrier. As code generation becomes highly automated, the critical bottleneck in modern web development has decisively shifted toward automated web testing~\citep{le2025automated, ye2025ai}.

Despite this shift, current research struggles to provide sufficient evaluation frameworks for this emerging paradigm. Existing efforts often emerge as byproducts of web generation tasks. Research that reconstructs web designs from UI screenshots assess generation quality through static visual similarity~\citep{si2025design2code, xiao2025designbench} or isolated interactive components~\citep{xiao2024interaction2code}. A more realistic approach synthesizes target webpages from text instructions and employs computer-use agents (CUAs) to simulate human behavior for dynamic functional evaluation~\citep{lu2025webgenbench,zhang2025artifactsbench}. Nevertheless, these methods heavily rely on predefined, rigid checklists, which constrain their utility in open-ended development environments. Furthermore, the unreliability of CUAs in complex web environments~\citep{zhou2024webarena,xie2024osworld} introduces execution bias into the functional evaluation process. On the other hand, recent testing application testing methods are similarly constrained by human-written checklists. Crucially, existing works overlook a fundamental aspect of software quality: potential logical constraints, \textit{e.g.}, in a meeting room reservation system, the same room cannot be double-booked within an overlapping time slot.

To address these gaps, we introduce \textbf{\dataset}, a benchmark for end-to-end long-horizon automated web testing, requiring agents to generate checklist and perform defect detection. To ground evaluation in realistic AI-driven web development scenarios, we synthesize 100 web applications spanning seven categories. We evaluate CUA performance along four quality dimensions: \textit{functionality}, \textit{constraint}, \textit{interaction}, and \textit{content}. Beyond standard functional verification and content relevance, \dataset also emphasis on latent logical constraints, a critical yet overlooked dimension in existing works. To evaluate LLM performance on \dataset, we design \textbf{\agent}, a two-stage framework for end-to-end web testing comprising a \textit{Checklist Generation Agent} and a \textit{Defect Detection Agent}. Building on this framework, we conduct a comprehensive evaluation of several powerful LLMs. The results are sobering: all evaluated models score below $30\%$ F1 on \dataset, revealing a substantial gap between current CUA capabilities and the demands of industrial-grade web testing deployment. In summary, our contributions are:

\begin{itemize}[leftmargin=*, nolistsep, noitemsep]
\item We introduce \dataset, a meticulously annotated web testing benchmark that simulates end-to-end automated testing scenarios for modern AI-driven web development, accompanied by a suite of evaluation metrics.
\item We design \agent, a baseline framework for end-to-end automated web testing built upon the \dataset evaluation protocol.
\item We conduct a comprehensive evaluation of several popular LLMs, uncovering their strengths and weaknesses across various dimensions. Hopefully, these findings provide solid guidance for advancing automated web testing in the era of vibe coding.
\end{itemize}

\section{Related Work}
\label{sec:related_work}

\begin{figure*}[t]
\centering
\includegraphics[width=\linewidth]{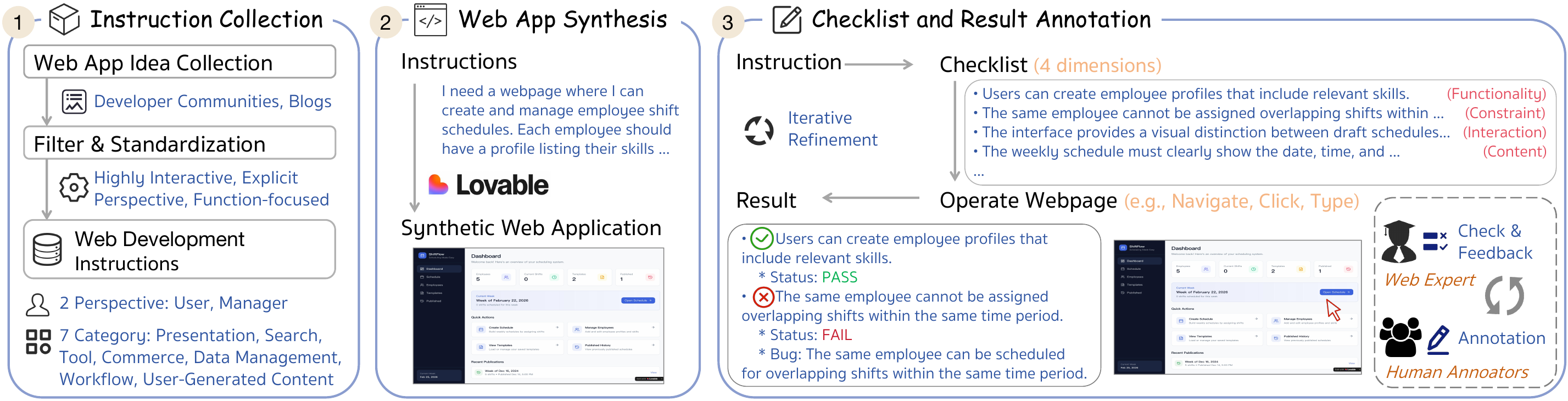}
\caption{Overview of \dataset construction.}
\label{fig:data_construction}
\vspace{-0.7em}
\end{figure*}

\textbf{Computer-Use Agents.} 
Beyond pure code generation, CUAs have emerged as a tool for interacting with web environments. Through screenshots~\citep{lin2025showui, nayak2025uivision, wang2025uitars2} or accessible code (\textit{e.g.}, DOM trees)~\citep{deng2023mind2web, chezelles2025browsergym}, CUAs can execute multi-step tasks involving navigation, clicking, and typing. Current benchmarks~\citep{zhou2024webarena, xie2024osworld, miyai2025webchorearena} primarily focus on operational success in realistic web scenarios, such as online shopping and discussion forums. While CUAs have demonstrated potential in general web activities, their effectiveness in software quality assurance, particularly defect detection, remains insufficiently explored.

\noindent\textbf{Web Testing Benchmarks.} Driven by the remarkable capabilities of CUAs and GUI agents in web task automation, recent works have introduced benchmarks to evaluate automated application testing. GTArena~\citep{zhao2024gtarena} proposes a comprehensive evaluation framework for automated GUI testing, decomposing the process into test intention generation, test task execution, and GUI defect detection. However, its defect detection is limited to atomic-level bugs inferred from single-step before-and-after state comparisons. GUITestBench~\citep{gao2026guitester} focuses on exploratory GUI defect discovery. \citet{chevrot2025planata} investigates the feasibility of adapting automated web agents as automated test agents, introducing a benchmark to assess their ability to execute human-written test cases. \citet{teoh2026webtestpilot} advances this further by additionally examining whether agents can accurately determine defect verdicts. Despite this progress, these works rely on human-written checklists, which presents a barrier for non-expert developers who lack the expertise to write comprehensive test cases. Furthermore, none of them account for latent logical constraints. \dataset aims to bridge these gaps, targeting the realistic capability of CUAs in checklist-free end-to-end automated web testing, and filling a critical missing piece in the application development lifecycle enabled by human-AI collaborative units.

\section{\dataset}
\label{sec:method}

The advent of AI-driven web development enable users to build web applications via natural language, yet introduces a new requirement, where novice creators lack the software engineering expertise to verify whether functions are reliably implemented. To systematically evaluate testing agents within this paradigm, we introduce \dataset.
In this section, we define the end-to-end automated web testing task, followed by a detailed description of our data construction and evaluation protocol.

\subsection{Task Definition}
\label{subsec:task_definition}

Given a user instruction and a corresponding web application, the target is to evaluate the end-to-end automated web testing capability of computer-use agents. Specifically, each agent is required to complete a two-stage task: 
\textbf{(1) Checklist Generation.} The agent decomposes the instruction into a set of verifiable test items, $\mathcal{T}_{\text{pred}} = \{p_1, \ldots, p_m\}$, where each test item $p_j$ describes a concrete behavior or property of the web application that can be validated through interaction. 
\textbf{(2) Defect Detection.} The agent plans and executes all test items by simulating human interactions (\textit{e.g.}, clicking, typing) within the web environment, and determines an execution status $s_j \in \{\text{Pass}, \text{Fail}\}$ for each item. 
The final output of the agent is a set of predicted test item-result pairs $\mathcal{P} = \{(p_j, s_j)\}_{j=1}^m$, which jointly represent the evaluation of the web application.

\subsection{Data Construction}
\label{subsec:data_construction}

\begin{figure*}[t]
\centering
\includegraphics[width=\linewidth]{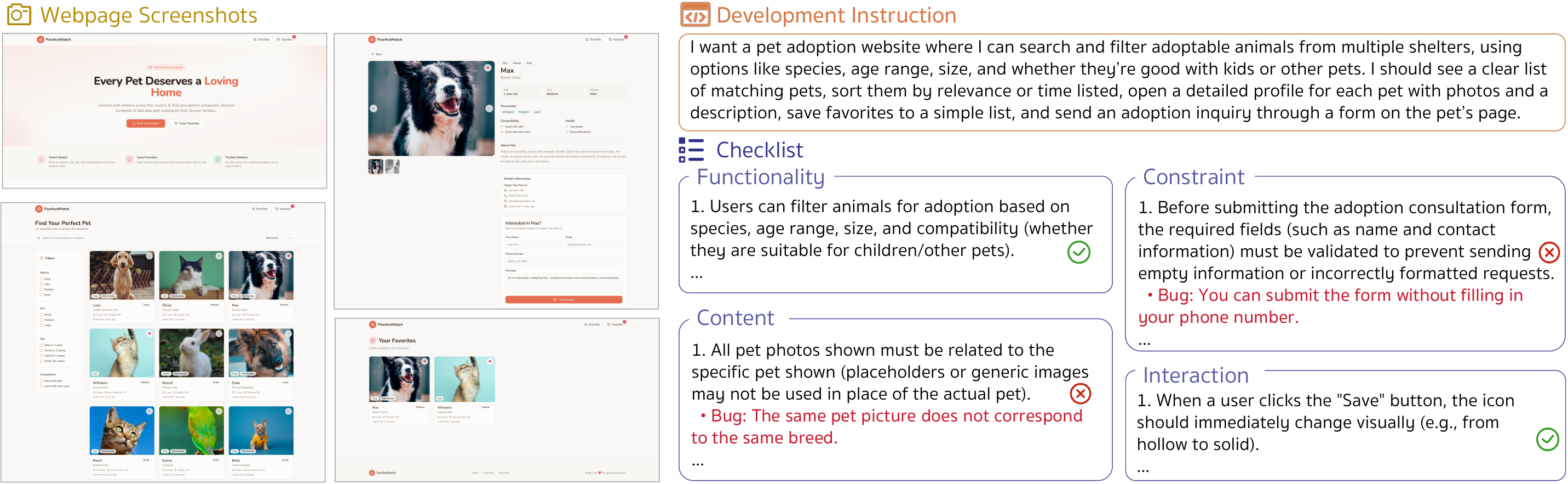}
\caption{An illustrative example from \dataset. The web application is built from a given development instruction, typically comprising multiple pages with rich interactive functionality. The benchmark provides a gold checklist across four quality dimensions: Functionality, Constraint, Interaction, and Content. Each test case is annotated with a binary Pass/Fail verdict.}
\label{fig:data_example}
\vspace{-0.7em}
\end{figure*}

Figure~\ref{fig:data_construction} illustrates the workflow for \dataset construction, primarily consisting of instruction collection, web application synthesis, and checklist and result annotation.

\noindent\textbf{Instruction Collection.} 
To ensure the diversity and practical relevance of our benchmark, we initially collect 451 trending web application ideas from diverse developer communities and blogs. We eliminate highly similar duplicates and filter these raw ideas to isolate those aligned with our evaluation objectives, prioritizing attributes such as interactivity, practicality, and testability. 

These ideas range from brief design sketches to overly verbose descriptions. Sketches are often too vague, while detailed descriptions include irrelevant content such as background narratives, references to commercial websites, and advertisements. We therefore rewrite each idea into a well-formed instruction using \texttt{GPT-5.1}. The model assigns each idea a specific perspective (\textit{i.e.}, \textit{User} or \textit{Administrator}) to disambiguate permission logic, categorizes it into one of seven functional themes (\textit{i.e.}, \textit{Presentation}, \textit{Search}, \textit{Tool}, \textit{Commerce}, \textit{Data Management}, \textit{Workflow}, and \textit{User-Generated Content}), and produces a formal development instruction. Subsequently, human annotators review and refine these instructions to ensure quality and standardization. Finally, we sample 100 instructions and maintain a balanced distribution across all categories.

\noindent\textbf{Web Application Synthesis.} 
Evaluating web testing agents requires environments that are ecologically valid and contain diverse and non-trivial defects. However, standard web resources often present limitations. Commercial websites are typically well tested and continuously updated, which makes them unsuitable as benchmark samples. Open-source projects often feature simple designs with shallow structures or limited interactive functionality, and therefore fail to reflect realistic user interactions. To bridge this gap, we utilize \texttt{Lovable.dev}\footnote{https://lovable.dev/}, an AI-powered web development platform that generates complete websites from user instructions, to synthesize web application projects. Through this process, we obtain an initial web application for each instruction, providing a realistic webpages for defect detection.

\noindent\textbf{Gold Checklist and Result Annotation.}
Given an development instruction and its application, human annotators construct a testable checklist. Inspired by software quality and evaluation standards (\textit{e.g.}, ISO/IEC 25010~\citep{isoiec25010}), and adapting them to the features of our benchmark, we categorize test cases into four dimensions: \textit{Functionality}, \textit{Constraint}, \textit{Interaction}, and \textit{Content}. Beyond isolated atomic interaction tests, we emphasize long-horizon interactions and latent logical constraints. For example, in a ``meeting room booking'' system, a functional test item would not merely be ``click a button'', but a complete sequence: ``select a specific room, complete the reservation form, and verify the successful status update.'' Additionally, potential logical checks are also necessary, such as ``the same meeting room cannot be reserved by two different users for the overlapping time slots.''

Annotators derive a checklist from the instruction, and then interact with the actual application to align test cases with the implemented components and interaction flows. Finally, annotators execute the checklist by interacting with the website and document the Pass/Fail status of each item and provide concise bug reports for failures.

\noindent\textbf{Iterative Refinement.}
Initial synthesis often produces applications with few defects, which limits the effectiveness in discriminatively evaluating web testing capabilities. To address this, annotators perform a iterative refinement. This involves revising the instruction for re-generate app or continuing the conversation with \texttt{lovable.dev} to add new features. Throughout this iteration, the checklist and results are updated synchronously until the samples contain sufficient defects for evaluation. 

\noindent\textbf{Quality Control.}
To ensure the quality of the dataset, all annotators undergo related training and conduct cross-validation during the annotation process. Finally, a senior annotation leader (non-authors) performs a final scan of the entire dataset, providing feedback and guiding annotators to optimize the annotations. This process ensures the standardization and high quality of the annotations.

\subsection{Evaluation Metrics}
\label{subsec:metrics}

Let $\mathcal{G} = \{(g_i, r_i)\}_{i=1}^n$ denote the gold outcome set, where $g_i$ represents a gold test item and $r_i \in \{\text{Pass, Fail}\}$ is its ground-truth status. To quantify performance, we compare the agent's outcome set $\mathcal{P}$ against an annotated gold outcome set $\mathcal{G}$. Specifically, for a generated checklist $\mathcal{T}_{\text{pred}} = \{p_1, \ldots, p_m\}$ and a gold checklist $\mathcal{T}_{\text{gold}} = \{g_1, \ldots, g_n\}$, each predicted item $p_j$ is mapped to $g_i$ if they are semantically equivalent in testing intent. We define a matching function $\phi: \mathcal{T}_{\text{pred}} \to \mathcal{T}_{\text{gold}} \cup \{\varnothing\}$, where $\phi(p_j) = g_i$ indicates that $p_j$ is matched to $g_i$. We define the matched set of a gold item $g_i$ as $\Phi(g_i) = \{\, p_j \in \mathcal{T}_{\text{pred}} \mid \phi(p_j) = g_i \,\}$. We employ \texttt{Qwen3.5-27B} as a semantic judge to perform this matching step.

\noindent\textbf{Coverage.} Coverage measures the completeness of the generated checklist, focusing on whether all necessary test items have been listed. It is defined as the proportion of gold test items successfully covered by the agent’s predictions:

\vspace{-1em}
\begin{equation}
    \text{Coverage} = \frac{\bigl|\{\, g_i \in \mathcal{T}_{\text{gold}} \mid \Phi(g_i) \neq \varnothing \,\}\bigr|}{|\mathcal{T}_{\text{gold}}|}
\end{equation}
\vspace{-0.7em}

\noindent\textbf{Defect Detection Performance.} We treat defect detection as a binary classification task where ``Fail'' (a real defect) is the positive class. For each gold item $g_i$, we aggregate its matched predictions into a single verdict:

\vspace{-1em}
\begin{equation}
D(g_i) = 
\begin{cases} 
\text{Fail}, & \!\!\!\text{if } \exists\, p_j \in \Phi(g_i),\; s_j = \text{Fail} \\ 
\text{Pass}, & \!\!\!\text{otherwise} \end{cases}
\end{equation}
\vspace{-0.7em}

where $D(g_i) = \text{Pass}$ when $\Phi(g_i) = \varnothing$. The confusion matrix entries are formalized as follows:

\begin{itemize}[leftmargin=*, nolistsep, noitemsep]

\item \textit{True Positive (TP):} The agent correctly identifies a ground-truth bug.

\vspace{-1em}
\begin{equation}
\text{TP} = \sum_{i=1}^{n} \mathbb{1}(r_i = \text{Fail} \land D(g_i) = \text{Fail})
\end{equation}
\vspace{-0.7em}

\item \textit{False Positive (FP):} The agent reports a defect for a gold item that is actually correct.

\vspace{-1em}
\begin{equation}
\text{FP} = \sum_{i=1}^{n} \mathbb{1}(r_i = \text{Pass} \land D(g_i) = \text{Fail})
\end{equation}
\vspace{-0.7em}

\item \textit{False Negative (FN):} The agent fails to detect a real bug, either due to misjudgment during detection or omission in checklist generation.

\vspace{-1em}
\begin{equation}
\text{FN} = \sum_{i=1}^{n} \mathbb{1}(r_i = \text{Fail} \land D(g_i) = \text{Pass})
\end{equation}
\vspace{-0.7em}

\item \textit{True Negative (TN):} The agent correctly raises no alarm for a passing gold item, either by correctly judging it as ``Pass'' or by not covering the item.

\vspace{-1em}
\begin{equation}
\text{TN} = \sum_{i=1}^{n} \mathbb{1}(r_i = \text{Pass} \land D(g_i) = \text{Pass})
\end{equation}

\end{itemize}

Based on these formal definitions, we can calculate Precision ($P$), Recall ($R$), and F1 score:

{
\vspace{-1em}
\small
\begin{equation}
P = \frac{\text{TP}}{\text{TP} + \text{FP}}, \; R = \frac{\text{TP}}{\text{TP} + \text{FN}}, \; \text{F1} = \frac{2 \times P \times R}{P + R}
\end{equation}
}

\subsection{\agent}
\label{subsec:baseline_agent}

Given the current absence of an established framework for end-to-end automated web testing, we introduce \agent, a minimal yet functional baseline designed for this task. Specifically, \agent consists of two components: a checklist generation agent and a defect detection agent. This framework enables autonomous testing of a web application based on the user instruction and the web application, ultimately producing structured test results accompanied by bug reports.

\noindent\textbf{Checklist Generation Agent.} Given a development instruction $I$ used to synthesize the target web application, the checklist generation Agent $A_C$ is responsible for decomposing $I$ into a structured, executable test checklist $\mathcal{T}_{\text{pred}}$. Formally, we define a test item $p$ as a triple:

\vspace{-1em}
\begin{equation}
    p = (d,\ a,\ e ),
\end{equation}

where $d$ is a textual description of what is being tested, $a$ is the action to be performed on the web interface (\textit{e.g.}, clicking the ``Edit'' button), and $e$ is the expected outcome that determines the Pass/Fail verdict. The full checklist generated by $A_C$ is a typed collection of such triples:

\vspace{-1em}
\begin{equation}
    \mathcal{T}_{\text{pred}} = A_C(I) = \{\, p_j \}_{j=1}^{m},
\end{equation}

\begin{table*}[t]
\centering
\resizebox{\textwidth}{!}{
\begin{tabular}{l|cc|cc|cc|cc|cc|cccc}
\toprule
\multirow{2}[1]{*}{Model} & \multirow{2}[1]{*}{\#Turns} & \multirow{2}[1]{*}{\#Tokens} & \multicolumn{2}{c|}{Functionality} & \multicolumn{2}{c|}{Constraint} & \multicolumn{2}{c|}{Interaction} & \multicolumn{2}{c|}{Content} & \multicolumn{4}{c}{Overall} \\
\cmidrule(lr){4-5} \cmidrule(lr){6-7} \cmidrule(lr){8-9} \cmidrule(lr){10-11} \cmidrule(lr){12-15}
& & & Cov. & F1 & Cov. & F1 & Cov. & F1 & Cov. & F1 & Cov. & P & R & F1 \\
\midrule
\rowcolor{gray!10}\multicolumn{15}{c}{\textit{\textbf{Open-Source LLMs}}}\\
Minimax-M2.1      & 41.7 & 3.58M & 77.9 & 12.3 & 40.4 & 15.8 & 42.2 & 19.9 & 47.1 &  \second{7.7} & 60.1 & 22.3 & 14.6 & 15.2 \\
Qwen3-Coder-Next  & 63.4 & 6.24M & 77.6 & 14.1 & 48.3 & 23.8 & 42.7 & 11.4 & 35.9 &  4.3 & 60.4 & 27.8 & 15.8 & 17.3 \\ 
GLM-4.7           & 41.6 & 3.47M & 79.9 & 16.5 & 47.3 & 20.5 & 36.0 & 17.2 & 48.9 &  4.3 & 61.1 & 26.7 & 16.6 & 18.1 \\
GLM-5             & 41.3 & 3.71M & 79.7 & 11.9 & 50.1 & 26.9 & 41.4 & 20.9 & 50.6 &  3.4 & 63.1 & 30.4 & 15.6 & 19.0 \\
Step-3.5-Flash    & 57.0 & 3.37M & 79.8 & 20.1 & \best{53.6} & \second{27.9} & 48.5 & 21.2 & \best{60.6} &  2.6 & \best{66.0} & \second{34.6} & 20.8 & 23.4 \\
MiMo-V2-Flash     & 59.8 & 7.26M & 80.0 & 21.9 & 48.7 & \best{29.2} & 48.0 & 20.3 & 50.3 &  7.3 & 63.5 & \best{34.8} & 24.6 & \second{25.1} \\
\rowcolor{gray!10}\multicolumn{15}{c}{\textit{\textbf{Closed-Source LLMs}}}\\
Claude Opus 4.5   & 42.9 & 2.60M & \best{83.3} & 18.8 & 50.3 & 21.2 & 40.8 & 14.7 & 42.1 &  6.8 & 63.2 & 33.0 & 16.5 & 20.2 \\
Claude Sonnet 4.5 & 37.6 & 1.90M & \second{81.0} & 22.2 & 47.7 & 22.5 & 46.7 & 19.9 & 51.6 &  1.7 & \second{63.7} & 32.1 & 19.7 & 21.9 \\
GPT-5.2           & 69.5 & 7.43M & 76.9 & \second{25.3} & \second{51.9} & 21.5 & 43.0 & \best{23.2} & 46.1 &  6.2 & 61.0 & 24.7 & \second{25.2} & 22.9 \\
GPT-5.1           & 30.3 & 0.87M & 76.4 & \best{30.9} & 51.2 & 26.9 & \best{49.7} & \second{22.0} & \second{57.5} & \best{15.3} & 63.1 & 25.8 & \best{33.3} & \best{26.4} \\
\bottomrule
\end{tabular}%
}
\caption{
Web testing performance of representative LLM on \dataset under the \agent framework. 
We report the Coverage metric (Cov.) for checklist generation, and the average number of iteration turns (\#Turns), average context tokens (\#Tokens) per instance, and Precision/Recall/F1 metrics for defect detection. Results in \best{bold} and \second{underline} denote the best and second-best performances.
}
\label{tab:main_results_overall}
\vspace{-0.5em}
\end{table*}

\noindent\textbf{Defect Detection Agent.} Given the development instruction $I$, the generated checklist $\mathcal{T}_{\text{pred}}$, and the target web application $W$, the defect detection agent $\text{A}_D$ is responsible for executing each test item $p_j \in \mathcal{T}_{\text{pred}}$ by simulating human interactions with the webpages, and determining a binary execution status for each item. Formally, $\mathcal{A}_D$ maps each test item together with the instruction and the web application to a verdict:

\vspace{-1em}
\begin{equation}
    s_j = A_D(p_j,\ I,\ W) \in \{\text{Pass},\ \text{Fail}\},
\end{equation}

For each failed item, $\mathcal{A}_D$ additionally produces a bug report $b_j$ describing the observed erroneous behavior. The final output of \agent is thus a set of predicted outcome triples:

\vspace{-1em}
\begin{equation}
    \mathcal{S} = A_D(\mathcal{T}_{\text{pred}},I,W) = \bigl\{ \bigl(p_j,s_j\ [,b_j]\bigr) \bigr\}_{j=1}^{m},
\end{equation}

\section{Experiments}
\label{sec:experiments}

\subsection{Settings}
\label{subsec:settings}

We build \agent on Claude Code as the basic agent framework. Specifically, we adopt the Claude Agent SDK\footnote{\url{https://github.com/anthropics/claude-agent-sdk-python}} to provide a unified evaluation interface across different LLMs. To interact with the web application, we leverage Playwright MCP\footnote{\url{https://github.com/microsoft/playwright-mcp}}, a Model Context Protocol server that provides browser automation capabilities (\textit{e.g.}, navigation, clicking, typing) to the agent. 

We evaluate several closed-source models and open-source models on \agent, including: Claude Opus/Sonnet 4.5~\cite{anthropic2025opus4.5,anthropic2025sonnet4.5}, GPT-5.2/5.1~\cite{openai2025gpt5.2,openai2025gpt5.1}, GLM-5/4.7~\cite{zeng2026glm5, zeng2025glm4.5}, Step-3.5-Flash~\cite{huang2026step3.5flash}, Qwen3-Coder-Next~\cite{qwen_qwen3_coder_next_tech_report}, MiMo-V2-Flash~\cite{xiao2026mimov2flash}, and Minimax-M2.1~\cite{minimax2025}. More details are available in Appendix \ref{appsubsec:more_settings}.

\subsection{Main Results}
\label{subsec:main_results}

\paragraph{Overall Performance.} Table~\ref{tab:main_results_overall} demonstrates the performance of the tested LLMs on end-to-end automated web testing. Despite advanced LLMs have made progress in agentic coding tasks and general web browsing tasks, they still exhibit notable performance bottlenecks when confronted with highly dynamic web testing scenarios involving complex interactions and constraints. All evaluated models failed to achieve an F1 score exceeding $30\%$. Among all evaluated models, GPT-5.1 attains the highest F1 score of 26.4\%, a standing driven primarily by its leading recall of 33.3\%. MiMo-V2-Flash ranks second with an F1 of 25.1\%, achieving the highest precision (34.8\%). 

\noindent\textbf{Insufficient Test Completeness.} 
Reliable defect detection relies on a comprehensive test checklist. However, across all tested models, their coverage consistently remains below 70\%. This indicates that a substantial portion of test items are excluded from subsequent detection and introduces false negatives. Even the two strongest models, GPT-5.1 and MiMo-V2-Flash, omit at least one-third of the test cases. This reflects a fundamental limitation in how LLMs decompose development instructions into verifiable test cases, particularly for implicit cases not explicitly stated in the instructions.

\begin{table*}[t]
\centering
\resizebox{\textwidth}{!}{
\begin{tabular}{l|ccccccc}
\toprule
Model & Presentation & Search & Tool & Commerce & DM & Workflow & UGC \\
\midrule
Minimax M2.1 & 25.6/10.3/14.3 & 18.9/6.9/9.4 & 12.3/12.6/11.7 & 10.9/10.1/8.6 & 36.4/23.4/25.3 & 18.4/12.3/12.8 & \best{27.2}/20.3/17.8 \\ 
Qwen3 Coder Next & 16.4/16.4/14.3 & 21.5/11.1/13.7 & 33.8/12.2/16.6 & 26.8/16.8/19.1 & 42.6/20.1/23.2 & 18.4/15.8/15.0 & \second{27.0}/15.9/16.4 \\ 
GLM 4.7 & 22.4/12.7/15.6 & 19.4/26.5/20.0 & 17.8/7.8/9.8 & \second{35.9}/19.1/24.1 & 41.8/22.1/24.7 & 22.6/15.7/16.3 & 21.7/16.2/16.3 \\ 
GLM 5 & 30.8/16.7/21.1 & 11.1/4.6/6.5 & 24.7/11.4/14.0 & 33.6/15.4/19.6 & 38.0/20.7/24.6 & 40.1/23.3/26.8 & 23.6/9.9/12.7 \\ 
Step 3.5 Flash & 38.6/13.2/18.8 & \second{24.1}/13.7/17.1 & \second{37.2}/22.2/25.3 & 34.3/\second{22.5}/\second{25.6} & \best{47.9}/\second{33.8}/\best{35.6} & 27.2/19.2/19.0 & 24.3/12.0/14.6 \\ 
MiMo-V2-Flash & 23.7/14.2/16.5 & \best{26.6}/\best{34.1}/\best{26.0} & \best{42.1}/28.1/\second{30.7} & 35.5/20.2/21.6 & \second{44.1}/33.2/\second{34.1} & \best{44.1}/24.0/28.4 & 13.8/16.1/10.6 \\ 
Claude Opus 4.5 & \second{41.2}/\second{25.0}/\best{30.5} & 7.4/4.6/5.6 & 25.5/15.7/16.2 & \best{55.1}/16.2/24.2 & 28.2/13.4/17.0 & \second{42.8}/24.5/\second{29.0} & 23.6/11.0/13.7 \\ 
Claude Sonnet 4.5 & \best{43.5}/19.1/23.8 & 21.1/20.9/19.7 & 32.6/22.4/24.6 & 22.7/13.0/14.0 & 43.8/23.4/26.9 & 31.4/22.8/24.7 & 20.1/12.6/14.3 \\ 
GPT 5.2 & 33.7/19.3/22.9 & 13.9/13.0/13.2 & 31.0/\best{35.1}/\best{31.9} & 20.8/19.9/18.2 & 28.4/24.4/23.9 & 21.8/\second{28.8}/23.6 & 16.6/\best{29.0}/\best{19.9} \\ 
GPT 5.1 & 27.6/\best{27.6}/\second{25.7} & 23.1/\second{31.3}/\second{23.4} & 25.0/\second{33.4}/26.4 & 28.7/\best{28.8}/\best{27.2} & 29.2/\best{39.9}/29.4 & 28.0/\best{38.2}/\best{29.5} & 15.5/\second{28.5}/\second{19.5} \\ 
\midrule
Avg. F1 ($\pm$Std.) & 20.4$_{\pm \text{5.4}}$ & 15.5$_{\pm \text{7.0}}$ & 20.9$_{\pm \text{7.8}}$ & 20.2$_{\pm \text{5.7}}$ & 26.5$_{\pm \text{5.4}}$ & 22.5$_{\pm \text{6.3}}$ & 15.6$_{\pm \text{3.6}}$ \\
\bottomrule
\end{tabular}
}
\caption{Category-wise evaluation results on \dataset. Each cell contains ``Precision/Recall/F1''. Results in \best{bold} and \second{underline} denote the best and second-best performances.}
\label{tab:main_results_by_category}
\vspace{-0.7em}
\end{table*}

\noindent\textbf{Detection Bottleneck.} 
CUAs exhibit severe deficiencies in web defect detection, taking in two primary failure modes. The first is a high false-positive rate. Most models achieve precision around 30\%, indicating that CUAs often misclassify benign behaviors as defects. For instance, transient UI rendering delays and asynchronous state updates are easily mistaken as functional failures, reflecting insufficient model understanding of dynamic web behavior. On the other hand, the low recall poses a greater risk. Most models fail to exceed a 25\% recall, meaning the major real defects remain undetected. Beyond limited defect cognition, these false-negatives are partly from a default-correctness bias, where models default to a pass judgment when no explicit evidence is observed. Additionally, we observe a strategic divergence in CUA behavior: they either employ an aggressive detection strategy that favors recall at the cost of precision (\textit{e.g.}, GPT-5.1 achieves 33.3\% recall but only 25.8\% precision) or adopt a conservative one that prioritizes precision while overlooking numerous real defects (\textit{e.g.}, MiMo-V2-Flash achieves 34.8\% precision but only 24.6\% recall).

\noindent\textbf{Long-horizon Interaction Unreliability.} 
Completing a comprehensive web defect detection process typically requires dozens of interaction turns and millions of tokens. For example, Step-3.5-Flash requires an average of 57.0 turns and 3.37M tokens per sample. Such long-horizon tasks demand rigorous long-context memory and planning stability. Specifically, as the interaction history accumulates, models become increasingly susceptible to tracking failures, resulting in the loss of prior states or the execution of redundant operations.

\begin{table}[!htbp]
\centering
\resizebox{0.45\textwidth}{!}{
\begin{tabular}{lccc}
\toprule
Model & Kendall's $\tau$ & Spearman's $\rho$ & Pearson $r$ \\
\midrule
GPT-5-Mini  & 68.8 & 75.0 & 78.9 \\
GPT-5       & 76.6 & 82.7 & 85.7 \\
Qwen3.5-27B & \best{78.5} & \best{83.2} & \best{87.1} \\
\bottomrule
\end{tabular}%
}
\caption{Human judgment correlation scores for our automatic evaluation. All $p$-values $<0.05$.}
\label{tab:human_consistency}
\vspace{-0.3em}
\end{table}

\noindent\textbf{Correlation with Human Judgments.} To validate the effectiveness and robustness of our automatic evaluation pipeline, we perform a correlation analysis between the automated evaluation results and human judgments, reporting Kendall's $\tau$, Spearman's $\rho$, and Pearson $r$ correlation coefficients. As shown in Table~\ref{tab:human_consistency}, all judge models achieve strong correlation with human judgments and Qwen3.5-27B attains the best correlation. Moreover, as an open-source model, it is not subject to deprecation or silent updates over time, ensuring long-term reproducibility and stability. We therefore adopt Qwen3.5-27B as the judge model in this work. More details are available in Appendix \ref{appsubsec:human_consistency}.

\subsection{Fine-grained Analysis}
\label{subsec:fine_grained_analysis}

\begin{figure*}[t]
\centering
\includegraphics[width=\linewidth]{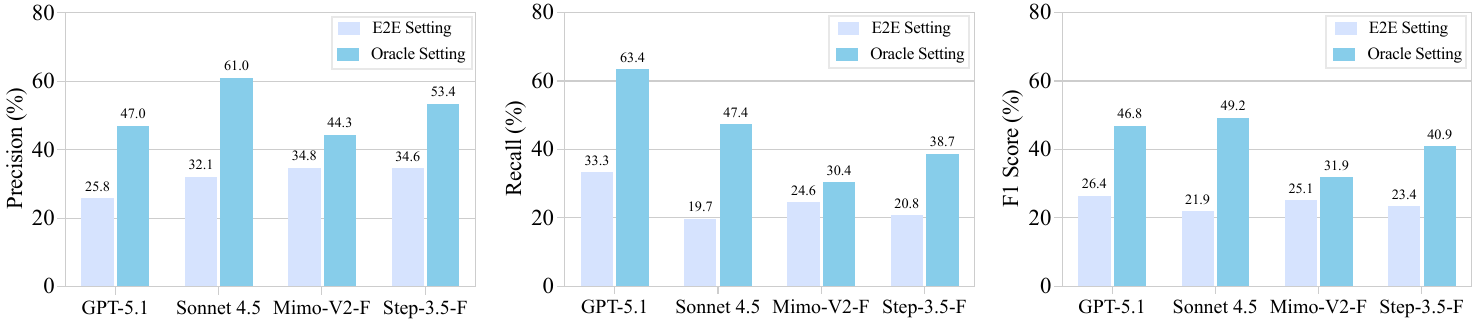}
\caption{Comparison of perfermance between the end-to-end (E2E) setting and the oracle setting for representative models. In the oracle setting, the gold checklist is directly provided to the defect detection agent, decoupling detection performance from checklist generation quality.}
\label{fig:oracle_results}
\vspace{-0.3em}
\end{figure*}

\textbf{Quality Dimensions.} 
While recent efforts have progressed toward comprehensive functional testing~\citep{lu2025webgenbench, zhang2025artifactsbench}, functional correctness alone is insufficient to capture overall application quality. To better understand where CUAs struggle in web testing, we report results across four quality dimensions.
As shown in Table~\ref{tab:main_results_overall}, \textit{Functionality} consistently achieves the high coverage across all models. This occurs because functional items are typically explicit and can be directly extracted from the development instruction. Conversely, coverage drops substantially for the other three types, which encompass implicit requirements demanding deeper reasoning. Despite lower coverage, \textit{Constraint} type achieves a relatively higher F1, revealing that once a constraint item is covered, defect detection becomes straightforward, as constraint violations typically produce clear, observable signals. However, \textit{Content} type performs worst by a marked margin. Error analysis reveals two primary failure modes: (1) models often fail to generate items targeting the actual defects, resulting in zero true positives and consequently zero F1 on many samples despite non-zero coverage. (2) models struggle to judge whether the displayed content is semantically aligned with the theme of the development instruction. These findings indicate that the primary bottleneck in checklist generation lies in the insufficient surfacing of implicit requirements, while content-level semantic alignment remains a formidable challenge.

\noindent\textbf{Category-wise Results.} Table~\ref{tab:main_results_by_category} reveals that performance varies across categories, driven by the nature of verification required. Models excel at \textit{Data Management}, where correctness reduces to tracking explicit, structurally observable state transitions (\textit{e.g.}, record creation/deletion/updates, or numerical changes). Performance degrades markedly in \textit{Search} and \textit{User-Generated Content}, exposing a semantic verification bottleneck: evaluating search relevance or dynamic content requires assessing abstract alignment with user intent, rather than observing discrete component states. Detailed analysis of each category are available in Appendix~\ref{appsubsec:detail_category_analysis}.

\subsection{Oracle Setting}
\label{subsec:fine_grained_analysis}

In the end-to-end setting, checklist generation and defect detection are executed sequentially, so final performance is jointly influenced by both sub-tasks. To isolate detection capability, we evaluate models in an oracle setting, where the defect detection agent receives the human-written gold checklist directly, thereby decoupling the detection phase from the upstream checklist generation.

As shown in Figure~\ref{fig:oracle_results}, all models achieve notable performance improvements in the oracle setting, confirming that test incompleteness is one of the primary bottlenecks in the end-to-end setting. The detection strategy preference observed earlier maintains consistent. GPT-5.1 remains aggressive (recall 63.4\%), while Claude Sonnet 4.5 remains conservative (precision 61.0\%). Notably, in oracle settings, closed-source models substantially outperform open-source models. This gap is not present in end-to-end setting, but becomes pronounced in oracle setting, indicating that for closed-source models, checklist generation is a more critical bottleneck in end-to-end testing. In comparison, Open-source models face challenges in both sub-tasks, and improving them requires advances in checklist coverage and detection reliability. 
More details and analysis are available in Appendix~\ref{appsubsec:detail_oracle}.

\begin{figure}[h]
\centering
\includegraphics[width=\linewidth]{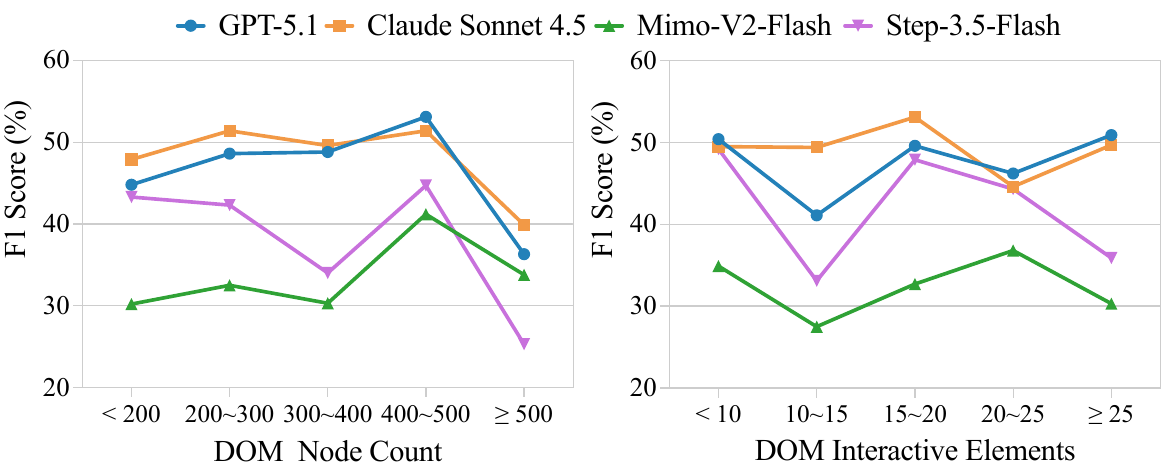}
\caption{Performance comparison across different web complexity in the oracle setting.}
\label{fig:web_complexity}
\end{figure}

\noindent\textbf{Web Complexity.} 
As shown in Figure~\ref{fig:web_complexity}, model performance generally degrades as web complexity increases, measured by DOM node count and the number of interactive elements per page. GPT-5.1 and Claude Sonnet 4.5 are comparatively robust, likely due to their more efficient interaction behavior (fewer turns, lower token usage), which reduces cascading errors in dense web environments. 

\section{Conclusion}
\label{sec:conclusion}

In this paper, we introduce \dataset, a challenging benchmark for end-to-end automated web testing without relying on human-written test cases, grounded in the context of AI-driven web development. Beyond standard functional testing, \dataset incorporates latent logical constraints, enabling a comprehensive assessment of application quality. We further propose \agent, a simple yet effective end-to-end web testing framework, and evaluate representative LLMs on \dataset. The results reveal a substantial gap between current CUA capabilities and the demands of industrial-grade deployment. Through comprehensive analyses, we identify the key pain points and multifaceted challenges. We hope \dataset serves as a foundation for future research in web testing during the vibe coding era, and encourages developing more reliable web testing agents.


\section*{Limitations}

Although \dataset provides fine-grained annotations across four quality dimensions, the construction process is inherently labor-intensive and demands domain expertise. Annotators must reason beyond explicit requirements to identify latent logical constraints. This specialized nature makes it costly to scale or replicate our methodology for building similar benchmarks. Second, our evaluation protocol also carries an inherent limitation. The scoring system relies on aligning predicted test items against the gold checklist. Thus, it cannot assess the validity of generated test cases that fall outside the gold checklist coverage, even if such cases reflect genuine quality concerns. Finally, while \dataset spans seven web application categories, it does not fully capture the diversity of real-world applications. Certain application types, such as ``games'' and ``complex multi-user real-time interactive systems'', are deliberately excluded, as they demand high-frequency capture and interpretation of highly dynamic content that exceeds the perceptual latency of both DOM-based and screenshot-based approaches, and thus lies beyond the current capability boundary of CUAs.
\section*{Ethics Policy}

The web development instructions collected for WebTestBench may contain background narratives or promotional content originating from real-world sources. Following our collection procedure, we remove or rewrite all such material, with human annotators verifying that no real-world entities remain. To support defect detection, we use \texttt{Lovable.dev} to populate each synthesized web application with sample data such as placeholder accounts and content. This data is not intended to represent actual individuals or organizations, and any resemblance is coincidental. Despite these precautions, the dataset may reflect latent social biases or stereotypes inherent in generative models, including those related to gender, race, age, and socioeconomic status. We therefore advise users to exercise caution when interpreting results or building upon this content.

\bibliography{cite, anthology}

\clearpage
\appendix

\section{\dataset}
\label{appsec:dataset}

\subsection{Dataset Statistics}
\label{subsec:dataset_statistics}

Table~\ref{tab:dataset_stats} presents the statistics of \dataset. The benchmark comprises 100 carefully curated development instructions spanning seven categories, each paired with an AI-generated web application that exhibits naturally occurring defects. The applications are structurally complex and highly interactive, reflecting the characteristics of real-world web development. On average, each instance is associated with 17.5 gold test items distributed across four dimensions: functionality, constraint, interaction, and content. The relevant descriptions and typical examples for the seven web application categories and four test case types are provided in Tables~\ref{tab:taxonomy_web_category} and \ref{tab:taxonomy_case_type}, respectively.

\subsection{Annotation}
\label{appsubsec:dataset_annotation}

\subsubsection{Annotation Document}
\label{appsubsubsec:anno_doc}

To standardize the annotation process, we prepare a detailed annotation document for human annotators covering four components: (1) a webpage category taxonomy, (2) a test case dimension taxonomy, (3) annotation content and procedures, and (4) expected outputs.

\noindent\textbf{Webpage Category Taxonomy.} We categorize web application development instructions into seven types. The annotation document provides annotators with the description, core functionalities, and typical examples of each category so that classification can be applied consistently. Detailed definitions are provided in Table~\ref{tab:taxonomy_web_category}.

\noindent\textbf{Test Case Dimension Taxonomy.} To enable fine-grained evaluation of CUA performance in end-to-end web testing, each test case is assigned to one of four dimensions. This design allows our experiments to analyze challenges arising from different aspects of the task. Notably, annotators should not merely extract explicitly mentioned requirements from development specifications. They must also apply additional understanding and reasoning to identify implicit test cases, such as potential corner cases. The annotation document provides descriptions, evaluation focus, and typical examples for each dimension. Detailed definitions are provided in Table~\ref{tab:taxonomy_case_type}.

\begin{table}[t]
\centering
\resizebox{0.4\textwidth}{!}{
\begin{tabular}{lr}
\toprule
\textbf{Statistic} & \textbf{Value} \\
\midrule
\multicolumn{2}{l}{\textit{Instructions \& Web Applications}} \\
\quad Total Samples                        & 100 \\
\quad \quad Presentation                   & 13 \\
\quad \quad Search                         &  9 \\
\quad \quad Tool                           & 17 \\
\quad \quad Commerce                       & 13 \\
\quad \quad Data Management                & 19 \\
\quad \quad Workflow                       & 17 \\
\quad \quad User-Generated Content         & 12 \\
\quad Avg. Instruction Tokens              & 127.8 \\
\quad Avg. Pages                           & 5.3 \\
\quad Avg. DOM Average Depth               & 9.4 \\
\quad Avg. DOM Node Count                  & 243.5 \\
\quad Avg. DOM Interactive Elements        & 18.9  \\
\quad Avg. DOM Unique Tag Count            & 31.1  \\
\midrule
\multicolumn{2}{l}{\textit{Test Cases}} \\
\quad Total items                          & 1750 \\
\quad \quad Functionality                  &  854 \\
\quad \quad Constraint                     &  398 \\
\quad \quad Interaction                    &  247 \\
\quad \quad Content                        &  251 \\
\quad Total Pass / Fail items              & 1302/448 \\
\quad \quad Functionality (Pass / Fail)    & 653/201 \\
\quad \quad Constraint   (Pass / Fail)     & 270/128 \\
\quad \quad Interaction  (Pass / Fail)     & 176/71 \\
\quad \quad Content      (Pass / Fail)     & 203/48 \\
\bottomrule
\end{tabular}
}
\caption{Overall statistics of \dataset.}
\label{tab:dataset_stats}
\vspace{-0.3em}
\end{table}

\noindent\textbf{Annotation Procedure.} Human annotators participate in the entire annotation pipeline. During instruction collection, annotators review and refine the development instructions rewritten by \texttt{GPT-5.1}, ensuring that the descriptions focus on application implementation and avoid references to real individuals, companies, or brands. They also verify and correct the category label assigned to each instruction. During web application synthesis, annotators use \texttt{Lovable.dev} to generate a web application from each instruction. During gold checklist and result annotation, annotators construct the required test cases by combining direct information from the instruction with additional reasoning about potential behaviors. Each test case is then assigned to one of the four dimensions. Annotators interact with the synthesized application to align test cases with the corresponding interface components and interaction flows, and perform defect detection for each item. This process produces a tuple consisting of the test case description, a verdict (Pass or Fail), and an optional bug report for failed items. If an instance contains fewer than three defects, annotators either continue interacting with \texttt{Lovable.dev} to introduce additional features or modify the instruction directly. The checklist and results are updated accordingly until the instance contains a sufficient number of defects, ensuring that each example is informative for defect detection evaluation. Finally, all annotation results are cross-validated among annotators.

\noindent\textbf{Expected Output.} Following this pipeline, each instance in the \dataset includes a development instruction, an web application containing naturally occurring defects, a manually written checklist, a Pass/Fail verdict for each test item, and an optional bug report for failed items.

\subsubsection{Human Annotators}
\label{appsubsubsec:annotators}

We employ crowdsourcing for data annotation. All annotators have prior experience evaluating the quality of AI-generated web applications in similar annotation tasks. Before the main annotation stage, candidates are required to study the annotation document and complete a trial annotation on three example samples. This stage serve both as training and as a qualification assessment. Only candidates who demonstrate high consistency, accuracy, and a clear understanding of the guidelines are selected for the main annotation task. A senior annotation lead, familiar with the task requirements and expected data quality, review all annotations and resolved disagreements identified during cross-validation.

\subsubsection{Reproducibility}
\label{appsubsubsec:reproducibility}

The code for each web application is available and can be deployed locally with a minimal setup command (\texttt{npm install \&\& npm run dev}) and does not rely on \texttt{Lovable.dev} during inference. This design ensures that the benchmark remains fully accessible and that experimental results can be reproduced independently of any future changes to the \texttt{Lovable.dev} platform.

\section{More Experimental Details and Results}
\label{appsec:more_experiments}

\subsection{More Experimental Settings}
\label{appsubsec:more_settings}

In this work, we employ Claude Code v2.1.25 and the Claude Agent SDK v0.1.0, with the maximum iteration turns limited to 150. To enable Claude Code to support non-Claude models, we integrate OpenRouter\footnote{\url{https://openrouter.ai/}} with Claude Code as a unified invocation interface. For the checklist generation agent, Playwright MCP is not employed, as the target task involves test case generation rather than web interaction. For the defect detection agent, we use Playwright MCP v0.0.41 with the viewport fixed at 1280$\times$720 during evaluation. The prompt templates for the two agents are shown in Figures~\ref{fig:prompt_checklist_generation} and \ref{fig:prompt_defect_detection}. When scoring, we set the temperature to 0.1 to ensure reproducibility. All other configurations follow the default settings.

While various LLMs can be deployed within Claude Code, the high complexity of the task presents substantial practical challenges. Models that fail to complete the testing process, such as inability to utilize Playwright MCP, output format errors, or unexpected termination, will be excluded from our evaluation. 

The matching between predicted and gold test items described in Section~\ref{subsec:metrics} is performed by \texttt{Qwen3.5-27B} as a semantic judge. The prompt template used for this step is provided in Figure~\ref{fig:prompt_match_item}.

\subsection{Correlation with Human Judgments}
\label{appsubsec:human_consistency}

As described in Section~\ref{subsec:metrics}, our evaluation protocol incorporates an LLM only for the relatively simple task of matching predicted test items to gold test items, rather than asking the LLM to directly assign scores. This design choice makes the evaluation more reliable compared to direct LLM-based scoring approaches.
To further validate the alignment between our automated evaluation protocol and human judgment, we conduct a correlation analysis on results aggregated from three models, including GLM-4.7, Step-3.5-Flash, and Claude Sonnet 4.5. Given the high cost of manual annotation, we randomly sample a subset of 20 instances with a balanced category distribution, comprising 352 gold test items and 1,144 predicted test items in total across the three models (GLM-4.7: 379, Step-3.5-Flash: 398, Claude Sonnet 4.5: 367). The per-category sample counts are 2/3/3/3/3/3/3 across the seven categories. Human annotators are asked to perform the item matching task manually, and the resulting evaluation metrics are compared against those produced by our automated pipeline. We report Kendall's $\tau$, Spearman's $\rho$, and Pearson $r$ measures. The results demonstrate that our automatic evaluation achieves strong alignment with human judgments across all correlation metrics.

\begin{table*}[h]
\centering

\resizebox{0.28\textwidth}{!}{
\begin{tabular}{p{3.8cm}r}
\toprule
\multicolumn{2}{c}{\textbf{GPT-5.1}} \\
\multicolumn{2}{r}{\small Avg. Playwright Calls: 23.23} \\
\midrule
\textbf{Tool} & \textbf{Calls} \\
\midrule
browser\_click & 14.69 \\ 
browser\_fill\_form & 2.14 \\ 
browser\_snapshot & 1.81 \\ 
browser\_wait\_for & 1.41 \\ 
browser\_navigate & 1.23 \\ 
browser\_evaluate & 0.68 \\ 
browser\_type & 0.55 \\ 
browser\_press\_key & 0.20 \\ 
\bottomrule
\end{tabular}
}
\hspace{1em}
\resizebox{0.28\textwidth}{!}{
\begin{tabular}{p{3.8cm}r}
\toprule
\multicolumn{2}{c}{\textbf{GPT-5.2}} \\
\multicolumn{2}{r}{\small Avg. Playwright Calls: 61.96} \\
\midrule
\textbf{Tool} & \textbf{Calls} \\
\midrule
browser\_click & 35.71  \\ 
browser\_snapshot & 6.47  \\ 
browser\_evaluate & 5.86  \\ 
browser\_fill\_form & 4.72  \\ 
browser\_navigate & 2.13  \\ 
browser\_wait\_for & 2.08  \\ 
browser\_type & 1.56  \\ 
browser\_press\_key & 1.36 \\ 
\bottomrule
\end{tabular}
}
\hspace{1em}
\resizebox{0.28\textwidth}{!}{
\begin{tabular}{p{3.8cm}r}
\toprule
\multicolumn{2}{c}{\textbf{Claude Sonnet 4.5}} \\
\multicolumn{2}{r}{\small Avg. Playwright Calls: 34.74} \\
\midrule
\textbf{Tool} & \textbf{Calls} \\
\midrule
browser\_click & 18.74  \\ 
browser\_snapshot & 4.73  \\ 
browser\_type & 3.39  \\ 
browser\_evaluate & 2.24  \\ 
browser\_navigate & 2.02  \\ 
browser\_fill\_form & 1.48  \\ 
browser\_wait\_for & 1.13  \\ 
browser\_press\_key & 0.58 \\ 
\bottomrule
\end{tabular}
}
\\
\vspace{0.5em}
\resizebox{0.28\textwidth}{!}{
\begin{tabular}{p{3.8cm}r}
\toprule
\multicolumn{2}{c}{\textbf{MiMo-V2-Flash}} \\
\multicolumn{2}{r}{\small Avg. Playwright Calls: 49.37} \\
\midrule
\textbf{Tool} & \textbf{Calls} \\
\midrule
browser\_click & 25.23  \\ 
browser\_evaluate & 8.29  \\ 
browser\_navigate & 4.00  \\ 
browser\_snapshot & 3.57  \\ 
browser\_fill\_form & 2.96  \\ 
browser\_type & 2.47  \\ 
browser\_wait\_for & 0.86  \\ 
browser\_console\_messages & 0.46 \\ 
\bottomrule
\end{tabular}
}
\hspace{1em}
\resizebox{0.28\textwidth}{!}{
\begin{tabular}{p{3.8cm}r}
\toprule
\multicolumn{2}{c}{\textbf{Step-3.5-Flash}} \\
\multicolumn{2}{r}{\small Avg. Playwright Calls: 48.83} \\
\midrule
\textbf{Tool} & \textbf{Calls} \\
\midrule
browser\_click & 26.01  \\ 
browser\_snapshot & 6.98  \\ 
browser\_type & 4.89  \\ 
browser\_evaluate & 3.30  \\ 
browser\_navigate & 2.59  \\ 
browser\_wait\_for & 1.78  \\ 
browser\_fill\_form & 1.47  \\ 
browser\_press\_key & 0.73 \\ 
\bottomrule
\end{tabular}
}
\hspace{1em}
\resizebox{0.28\textwidth}{!}{
\begin{tabular}{p{3.8cm}r}
\toprule
\multicolumn{2}{c}{\textbf{GLM-5}} \\
\multicolumn{2}{r}{\small Avg. Playwright Calls: 37.73} \\
\midrule
\textbf{Tool} & \textbf{Calls} \\
\midrule
browser\_click & 23.84  \\ 
browser\_snapshot & 4.33  \\ 
browser\_type & 3.90  \\ 
browser\_navigate & 1.74  \\ 
browser\_fill\_form & 1.63  \\ 
browser\_evaluate & 0.69  \\ 
browser\_press\_key & 0.62  \\ 
browser\_wait\_for & 0.37 \\ 
\bottomrule
\end{tabular}
}
\caption{Statistics of Playwright tool calls per instance for representative models.}
\label{tab:tool_call_stats}
\end{table*}

\subsection{Detailed Category-wise Analysis}
\label{appsubsec:detail_category_analysis}

As shown in Table~\ref{tab:detail_results_by_category}, we present detailed category-wise results, along with the cross-model mean and standard deviation for each category and metric, to reflect both the average performance and the variability across models. We further analyze these results in conjunction with the category definitions provided in Table~\ref{tab:taxonomy_web_category}.

\textit{Data Management} achieves the highest overall performance, with an average F1 score of 26.5\%. Applications in this category involve explicit state transitions, such as record creation, deletion, and numerical updates, where Pass/Fail judgments can be directly inferred from observable changes in the interface state before and after an operation. \textit{Workflow} applications share a similar state-transition structure but present additional challenges due to multi-step processes and the need for cross-stage context tracking.

\textit{Commerce} and \textit{Tool} applications occupy the middle tier, with average F1 scores of 20.2\% and 20.9\%, respectively. Commerce applications involve structured transaction flows that can generally be verified, though complex business logic, such as inventory constraints and payment validation, introduces additional reasoning requirements. Tool applications follow an input-processing-output paradigm, where correctness verification requires models to infer expected outputs based on specific inputs. Notably, \textit{Tool} shows the largest cross-model variance in F1 (7.8\%), suggesting that the reasoning demands in this category are unevenly handled across models.

\textit{Presentation} performance is dominated by closed-source models, with Claude Opus 4.5 achieving the highest F1 score of 30.5\%, while open-source models generally perform much worse. This gap suggests that evaluating content layout and semantic structure benefits disproportionately from stronger model capacity. \textit{Search} shows similar large cross-model variance (7.0\%), with performance ranging from 5.6\% for Claude Opus 4.5 to 26.0\% for MiMo-V2-Flash. Both categories require assessing semantic alignment between rendered content and user intent, rather than simply observing discrete component states, which presents a verification challenge that no model currently handles consistently.

\textit{User-Generated Content} has the worst performance, with an average F1 score of only 15.6\% and the smallest cross-model variance (S3.0\%), indicating that all models struggle equally in this domain. UGC applications involve unstructured, dynamically generated content, whose correctness cannot be determined from fixed component states, necessitating abstract semantic judgment that consistently exceeds current CUA capabilities.

Together, these disparities emphasize the fundamental distinction between operation-driven verification, where correctness is grounded in observable state transitions, and understanding-driven verification, where correctness relies on semantic alignment with user intent. Overcoming this gap presents a central challenge for the future development of automated web testing.

\subsection{Analysis of Tool Usage}
\label{appsubsec:analysis_tool_usage}

To better understand the interaction workflows of CUAs during automated web testing, we analyze the playwright tool usage of representative models, as shown in Table~\ref{tab:tool_call_stats}.

We observe substantial variation in interaction volume across models. GPT-5.2 produces the highest number of tool calls (61.96 per instance), indicating a preference for aggressive exploration through frequent web interactions. In contrast, GPT-5.1 uses fewer calls (23.23 per instance), suggesting a more efficiency-oriented strategy, likely supported by stronger planning capabilities.

Across all models, \texttt{browser\_click} is the most frequently used tool, accounting for approximately 50\% to 63\% of total calls. This is consistent with real-world web interaction patterns, where click-based actions constitute the most fundamental and pervasive form of engagement. In addition, notable differences emerge in how models leverage auxiliary tools. MiMo-V2-Flash and GPT-5.2 rely more heavily on \texttt{browser\_evaluate} (8.29 and 5.86 per instance respectively), indicating a tendency to inspect underlying page structures such as the DOM structure or JavaScript execution results. While this approach can enable more precise state verification, it also incurs considerable computational overhead, as reflected in the significantly higher context token consumption of both models (7.26M and 7.43M respectively). Notably, these models exhibit relatively low usage of \texttt{browser\_wait\_for}, suggesting that asynchronous behaviors are not adequately handled. This may cause models to misinterpret transient UI rendering delays and asynchronous state updates as genuine defects, contributing to elevated false-positive rates.

These findings highlight that future CUA systems would benefit from more deliberate interaction planning and more selective use of tools, both of which are important for improving testing reliability in dynamic web environments.

\begin{figure*}[t]
\centering
\includegraphics[width=0.9\linewidth]{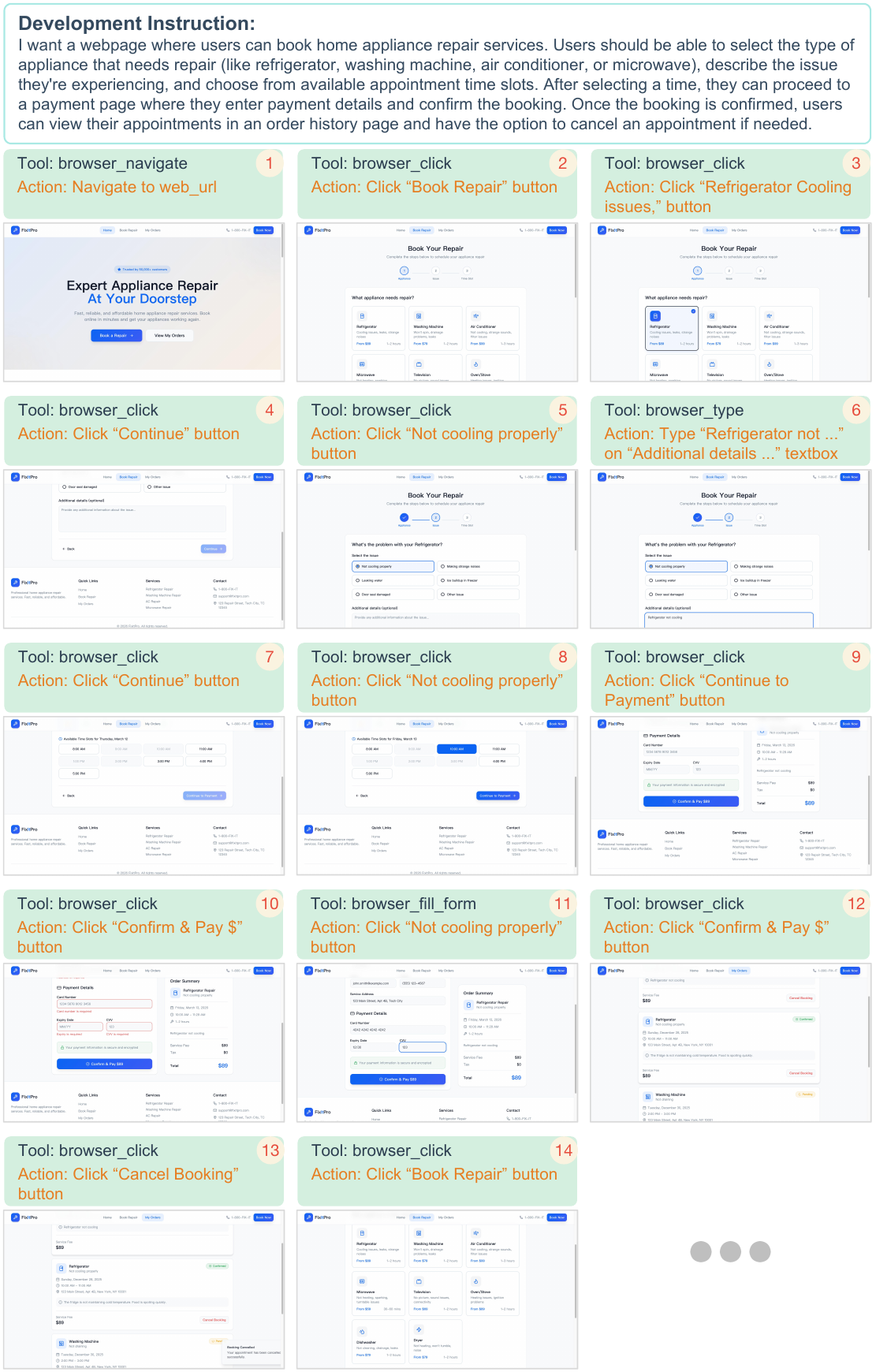}
\caption{An example illustrating how CUA interacts with a web application during defect detection.}
\label{fig:visual_interaction}
\end{figure*}

In addition, Figure~\ref{fig:visual_interaction} illustrates the web interaction process of CUAs during defect detection. We present the invoked Playwright tools, the corresponding actions, and the resulting webpage screenshots after each interaction. This visualization is based on the output of GPT-5.1.

\subsection{Detailed Results and Analysis in the Oracle Setting}
\label{appsubsec:detail_oracle}

Table~\ref{tab:detail_oracle_results} reports detailed results under the oracle setting, where the gold checklist is directly provided to the defect detection agent. Compared to the end-to-end setting, all models show clear improvements, indicating that incomplete checklists are the major bottleneck in end-to-end testing. In this setting, the gap between closed-source and open-source models also becomes more pronounced. Claude Sonnet 4.5 and GPT-5.1 achieve overall F1 scores of 49.2\% and 46.8\%, respectively, clearly outperforming MiMo-V2-Flash (31.9\%) and Step-3.5-Flash (40.9\%). This pattern indicates that checklist generation is the primary limiting factor for closed-source models in end-to-end setting, while open-source models still face difficulties in both checklist generation and detection.

Across all models, \textit{Constraint} consistently achieves the highest F1 scores, with Claude Sonnet 4.5 reaching 66.7\% and GPT-5.1 reaching 59.2\%. This is consistent with the nature of constraint violations, which often appear as explicit and observable state changes that are easy to verify once the relevant test item is given. In contrast, \textit{Functionality} remains difficult for open-source models even in the oracle setting, suggesting limitations in handling multi-step interactions and reasoning over execution processes. Notably, \textit{Content} shows the lowest performance across all models. This indicates that verifying semantic alignment between displayed content and development intent or test query remains challenging.

\begin{table*}[t]
\centering
\resizebox{\textwidth}{!}{
\begin{tabular}{p{2.8cm} p{13.6cm}}
\toprule
\textbf{Category} & \textbf{Details} \\
\midrule

\multirow{6}{=}{\textbf{Presentation}}
& \textit{Description:} Presents information in a clear, structured form, , enabling users to browse, read, and access content efficiently. \\
& \textit{Core Functionalities:} Content layout and typography, navigation structure, pagination and indexing, media rendering, basic search and filtering. \\
& \textit{Typical Examples:} Blog and article websites, personal homepages, corporate websites, documentation and wikis, portfolios, and news portals. \\
\midrule

\multirow{6}{=}{\textbf{Search}}
& \textit{Description:} Supports search, filtering, ranking, and recommendation over heterogeneous datasets, emphasizing information discovery and matching. \\
& \textit{Core Functionalities:} Search input and query syntax, filter and sort controls, result list presentation, recommendation-based content delivery, result detail pages. \\
& \textit{Typical Examples:} Job platforms, real estate listing websites, academic literature search engines, content recommendation systems. \\
\midrule

\multirow{7}{=}{\textbf{Tool}}
& \textit{Description:} Provides a structured input--process--output workflow to help users perform transformation, analysis, or generation tasks, typically emphasizing accuracy and clear feedback. \\
& \textit{Core Functionalities:} Input collection (forms), algorithmic or logical processing, result presentation and visualization, and editors for text, graphics, or tables. \\
& \textit{Typical Examples:} Data calculation and conversion tools, visualization generators, online document editors, format converters, simulators and configurators. \\
\midrule

\multirow{7}{=}{\textbf{Commerce}}
& \textit{Description:} Supports complete transaction flows, from product or service selection, often involving business logic related to inventory, pricing, orders, and payment verification. \\
& \textit{Core Functionalities:} Product and service browsing, shopping cart and checkout processes, payment and confirmation mechanisms, order management, inventory and pricing management, reservation and scheduling features. \\
& \textit{Typical Examples:} E-commerce websites, booking systems for hotels, flights, restaurants, or meeting rooms, subscription platforms, second-hand marketplaces. \\
\midrule

\multirow{7}{=}{\textbf{Data\\Management}}
& \textit{Description:} Provides management, monitoring, and operational capabilities over structured data, typically in internal or back-office settings, emphasizing organization, editing, and visualization. \\
& \textit{Core Functionalities:} Data tables with CRUD operations, access control mechanisms, dashboards and audit logs, statistical charts, and system configuration. \\
& \textit{Typical Examples:} User, order, and inventory management systems, enterprise back-office platforms, visual analytics dashboards. \\
\midrule

\multirow{6}{=}{\textbf{Workflow}}
& \textit{Description:} Organizes user tasks through predefined steps or stages, emphasizing process control, state management, approval flows, and cross-role collaboration. \\
& \textit{Core Functionalities:} Multi-step forms, tracking and state transitions, task and ticket management, approval workflows, content management modules, cross-role collaboration. \\
& \textit{Typical Examples:} Content management systems (CMS), project management systems, task and approval systems, complex registration or data collection pipelines. \\
\midrule

\multirow{6}{=}{\textbf{User-Generated\\Content}}
& \textit{Description:} Core content is created, edited, shared, and interacted with by users, emphasizing community dynamics, creation toolchains, and social networks. \\
& \textit{Core Functionalities:} Content publishing and editors, comments and reactions, user profiles and social graphs, content moderation, notifications and real-time updates. \\
& \textit{Typical Examples:} Online forums, commenting systems, social platforms, creative platforms for writing, images, or video content. \\
\bottomrule
\end{tabular}
}
\caption{Taxonomy of web application categories used in \dataset.}
\label{tab:taxonomy_web_category}
\end{table*}

\begin{table*}[t]
\centering
\resizebox{\textwidth}{!}{
\begin{tabular}{p{2.5cm} p{13cm}}
\toprule
\textbf{Dimension} & \textbf{Details} \\
\midrule

\multirow{8}{=}{\textbf{Functionality}}
& \textit{Description:} Focuses on whether the application can successfully complete the primary tasks and business flows specified in the development instruction under normal usage conditions. \\
& \textit{Evaluation Focus:} Successful completion of core operations (create, view, edit, delete, submit); correct responses to normal user requests (search, filter, navigate, form submission); coherent cross-module flows and reachable navigation. \\
& \textit{Typical Examples:} A user can submit a search query and view the returned results; a user can create a new project and see it reflected in the list. \\
\midrule

\multirow{9}{=}{\textbf{Constraint}}
& \textit{Description:} Focuses on whether the application is robust under abnormal or boundary usage, including input validation, error handling, and prevention of invalid or conflicting states. \\
& \textit{Evaluation Focus:} Effective validation of required fields with clear error messages; blocking of conflicting or illegal operations (e.g., double-booking, editing non-existent records); graceful handling of boundary conditions (empty lists, excessively long inputs, invalid formats); consistency between interface state and underlying data. \\
& \textit{Typical Examples:} An already-reserved meeting room cannot be booked again; submitting a form with empty required fields is blocked with a clear prompt. \\
\midrule

\multirow{9}{=}{\textbf{Interaction}}
& \textit{Description:} Focuses on the dynamic feedback and visual or state changes the application provides in response to user actions, reflecting the experiential quality of functionality rather than its correctness. \\
& \textit{Evaluation Focus:} Clear feedback after click, input, or toggle actions (e.g., button state changes, toast notifications, success or failure messages); smooth state transitions such as tab switching, expand/collapse, and toggle; visible and interpretable action outcomes (e.g., redirect or confirmation after submission). \\
& \textit{Typical Examples:} A success notification appears after a booking is created; hovering over a button triggers an appropriate visual change. \\
\midrule

\multirow{8}{=}{\textbf{Content}}
& \textit{Description:} Focuses on whether the information displayed by the application is accurate, complete, and semantically consistent with the development instruction, covering all front-end content elements including text, data, images, and icons. \\
& \textit{Evaluation Focus:} Thematic consistency of text, field labels; complete and well-formatted content in lists, detail pages, and cards; media assets (images, icons) rendered correctly and relevant to the theme. \\
& \textit{Typical Examples:} All images on an iPhone showcase page are iPhone-related; each blog card includes the title, summary, and date fields required by the instruction. \\
\bottomrule
\end{tabular}
}
\caption{Taxonomy of test case dimensions used in \dataset.}
\label{tab:taxonomy_case_type}
\vspace{+5em}
\end{table*}

\begin{table*}[t]
\centering

\resizebox{\textwidth}{!}{
\begin{tabular}{l|cccc|cccc|cccc|cccc}
\toprule
\multirow{2}[1]{*}{Model} & \multicolumn{4}{c|}{Functionality} & \multicolumn{4}{c|}{Constraint} & \multicolumn{4}{c|}{Interaction} & \multicolumn{4}{c}{Content} \\
\cmidrule(lr){2-5} \cmidrule(lr){6-9} \cmidrule(lr){10-13} \cmidrule(lr){14-17} \cmidrule(lr){12-15}
& Cov. & P & R & F1 & Cov. & P & R & F1 & Cov. & P & R & F1 & Cov. & P & R & F1  \\
\midrule
Minimax M2.1 & 77.9 & 15.9 & 14.0 & 12.3 & 40.4 & 19.9 & 14.4 & 15.8 & 42.2 & 22.6 & 20.3 & 19.9 & 47.1 & \second{8.6} & 7.3 & \second{7.7} \\ 
Qwen3 Coder Next & 77.6 & 17.5 & 14.9 & 14.1 & 48.3 & 27.5 & 23.1 & 23.8 & 42.7 & 12.4 & 12.8 & 11.4 & 35.9 & 5.1 & 3.9 & 4.3 \\ 
GLM 4.7 & 79.9 & 21.0 & 16.8 & 16.5 & 47.3 & 26.5 & 18.4 & 20.5 & 36.0 & 18.9 & 17.6 & 17.2 & 48.9 & 5.1 & 3.9 & 4.3 \\ 
GLM 5 & 79.7 & 16.8 & 11.0 & 11.9 & 50.1 & 32.4 & 26.5 & 26.9 & 41.4 & \second{24.2} & 20.3 & 20.9 & 50.6 & 5.1 & 2.6 & 3.4 \\ 
Step 3.5 Flash & 79.8 & \second{28.1} & 18.7 & 20.1 & \best{53.6} & \second{32.6} & 27.4 & \second{27.9} & \second{48.5} & \second{24.2} & 21.2 & 21.2 & \best{60.6} & 2.6 & 2.6 & 2.6 \\ 
MiMo-V2-Flash & 80.0 & 24.8 & 24.8 & 21.9 & 48.7 & \best{34.2} & \best{28.6} & \best{29.2} & 48.0 & 22.6 & 19.3 & 20.3 & 50.3 & 7.3 & \second{9.0} & 7.3 \\ 
Claude Opus 4.5 & \best{83.3} & 23.2 & 18.1 & 18.8 & 50.3 & 27.4 & 18.3 & 21.2 & 40.8 & 18.6 & 13.4 & 14.7 & 42.1 & 7.7 & 6.4 & 6.8 \\ 
Claude Sonnet 4.5 & \second{81.0} & 26.1 & 23.3 & 22.2 & 47.7 & 29.0 & 20.3 & 22.5 & 46.7 & 22.6 & 20.3 & 19.9 & 51.6 & 2.6 & 1.3 & 1.7 \\ 
GPT 5.2 & 76.9 & 25.4 & \second{31.9} & \second{25.3} & \second{51.9} & 23.1 & 23.1 & 21.5 & 43.0 & \best{27.5} & \second{21.9} & \best{23.2} & 46.1 & 7.7 & 5.8 & 6.2 \\ 
GPT 5.1 & 76.4 & \best{28.7} & \best{42.1} & \best{30.9} & 51.2 & 29.6 & \second{27.5} & 26.9 & \best{49.7} & 23.5 & \best{23.5} & \second{22.0} & \second{57.5} & \best{14.5} & \best{16.7} & \best{15.3} \\ 
\bottomrule
\end{tabular}
}

\caption{Detailed quality dimension-wise evaluation results on \dataset.}
\label{tab:detail_results_by_dimension}
\end{table*}

\begin{table*}[t]
\centering

\resizebox{\textwidth}{!}{
\begin{tabular}{l|cccc|cccc|cccc|cccc}
\toprule
\multirow{2}[1]{*}{Model} & \multicolumn{4}{c|}{Presentation} & \multicolumn{4}{c|}{Search} & \multicolumn{4}{c|}{Tool} & \multicolumn{4}{c}{Commerce} \\
\cmidrule(lr){2-5} \cmidrule(lr){6-9} \cmidrule(lr){10-13} \cmidrule(lr){14-17} \cmidrule(lr){12-15}
& Cov. & P & R & F1 & Cov. & P & R & F1 & Cov. & P & R & F1 & Cov. & P & R & F1  \\
\midrule
Minimax M2.1 & 56.9 & 25.6 & 10.3 & 14.3 & 52.0 & 18.9 & 6.9 & 9.4 & 58.7 & 15.2 & 13.7 & 13.4 & 57.1 & 10.9 & 10.1 & 8.6 \\
Qwen3 Coder Next & 62.5 & 16.4 & 16.4 & 14.3 & 62.7 & 21.5 & 11.1 & 13.7 & 56.1 & 33.8 & 12.2 & 16.6 & 66.1 & 26.8 & 16.8 & 19.1 \\
GLM 4.7 & 59.5 & 22.4 & 12.7 & 15.6 & 63.9 & 19.4 & 26.5 & 20.0 & \second{62.4} & 17.8 & 7.8 & 9.8 & 63.4 & \second{35.9} & 19.1 & 24.1 \\
GLM 5 & \second{64.5} & 30.8 & 16.7 & 21.1 & 61.6 & 11.1 & 4.6 & 6.5 & \best{64.0} & 24.7 & 11.4 & 14.0 & 64.9 & 33.6 & 15.4 & 19.6 \\
Step 3.5 Flash & \best{65.7} & 38.6 & 13.2 & 18.8 & \best{68.4} & \second{24.1} & 13.7 & 17.1 & 61.4 & \second{37.2} & 22.2 & 25.3 & \best{72.8} & 34.3 & \second{22.5} & \second{25.6} \\
MiMo-V2-Flash & 58.5 & 23.7 & 14.2 & 16.5 & \second{65.0} & \best{26.6} & \best{34.1} & \best{26.0} & 59.8 & \best{42.1} & 28.1 & \second{30.7} & 66.2 & 35.5 & 20.2 & 21.6 \\
Claude Opus 4.5 & 64.6 & \second{41.2} & \second{25.0} & \best{30.5} & 59.5 & 7.4 & 4.6 & 5.6 & 62.1 & 25.5 & 15.7 & 16.2 & 63.6 & \best{55.1} & 16.2 & 24.2 \\
Claude Sonnet 4.5 & 62.1 & \best{43.5} & 19.1 & 23.8 & 64.6 & 21.1 & 20.9 & 19.7 & 61.9 & 32.6 & 22.4 & 24.6 & 63.5 & 22.7 & 13.0 & 14.0 \\
GPT 5.2 & 53.4 & 33.7 & 19.3 & 22.9 & 59.4 & 13.9 & 13.0 & 13.2 & 60.7 & 31.0 & \best{35.1} & \best{31.9} & 61.5 & 20.8 & 19.9 & 18.2 \\
GPT 5.1 & 55.6 & 27.6 & \best{27.6} & \second{25.7} & 62.8 & 23.1 & \second{31.3} & \second{23.4} & 60.0 & 25.0 & \second{33.4} & 26.4 & \second{70.1} & 28.7 & \best{28.8} & \best{27.2} \\
\midrule
\textit{Avg.} & 60.3 & 30.4 & 17.5 & 20.4 & 62.0 & 18.7 & 16.7 & 15.5 & 60.7 & 28.5 & 20.2 & 20.9 & 64.9 & 30.4 & 18.2 & 20.2 \\ 
\textit{Std.} & 4.2 & 8.8 & 5.5 & 5.4 & 4.4 & 6.1 & 10.9 & 7.0 & 2.2 & 8.4 & 9.6 & 7.8 & 4.4 & 11.7 & 5.2 & 5.7 \\
\bottomrule
\end{tabular}
}

\vspace{5pt}

\resizebox{\textwidth}{!}{
\begin{tabular}{l|cccc|cccc|cccc|cccc}
\toprule
\multirow{2}[1]{*}{Model} & \multicolumn{4}{c|}{Data Management} & \multicolumn{4}{c|}{Workflow} & \multicolumn{4}{c|}{User-Generated Content} & \multicolumn{4}{c}{Overall} \\
\cmidrule(lr){2-5} \cmidrule(lr){6-9} \cmidrule(lr){10-13} \cmidrule(lr){14-17} \cmidrule(lr){12-15}
& Cov. & P & R & F1 & Cov. & P & R & F1 & Cov. & P & R & F1 & Cov. & P & R & F1  \\
\midrule
Minimax M2.1 & 63.8 & 36.4 & 23.4 & 25.3 & 65.6 & 18.4 & 12.3 & 12.8 & 60.9 & \best{27.2} & 20.3 & 17.8 & 60.1 & 22.3 & 14.6 & 15.2 \\
Qwen3 Coder Next & 59.3 & 42.6 & 20.1 & 23.2 & 59.4 & 18.4 & 15.8 & 15.0 & 59.4 & \second{27.0} & 15.9 & 16.4 & 60.4 & 27.8 & 15.8 & 17.3 \\
GLM 4.7 & 59.1 & 41.8 & 22.1 & 24.7 & 61.1 & 22.6 & 15.7 & 16.3 & 59.7 & 21.7 & 16.2 & 16.3 & 61.1 & 26.7 & 16.6 & 18.1 \\
GLM 5 & 61.3 & 38.0 & 20.7 & 24.6 & 65.5 & 40.1 & 23.3 & 26.8 & 58.8 & 23.6 & 9.9 & 12.7 & 63.1 & 30.4 & 15.6 & 19.0 \\
Step 3.5 Flash & \best{67.3} & \best{47.9} & \second{33.8} & \best{35.6} & 65.0 & 27.2 & 19.2 & 19.0 & \second{63.5} & 24.3 & 12.0 & 14.6 & \best{66.0} & 34.6 & 20.8 & 23.4 \\
MiMo-V2-Flash & 66.4 & \second{44.1} & 33.2 & \second{34.1} & 64.9 & \best{44.1} & 24.0 & 28.4 & \best{63.7} & 13.8 & 16.1 & 10.6 & 63.5 & \best{34.8} & 24.6 & \second{25.1} \\
Claude Opus 4.5 & 63.0 & 28.2 & 13.4 & 17.0 & \second{66.4} & \second{42.8} & 24.5 & \second{29.0} & 61.0 & 23.6 & 11.0 & 13.7 & 63.2 & 33.0 & 16.5 & 20.2 \\
Claude Sonnet 4.5 & 64.7 & 43.8 & 23.4 & 26.9 & \best{67.2} & 31.4 & 22.8 & 24.7 & 60.8 & 20.1 & 12.6 & 14.3 & \second{63.7} & 32.1 & 19.7 & 21.9 \\
GPT 5.2 & 62.4 & 28.4 & 24.4 & 23.9 & 65.3 & 21.8 & \second{28.8} & 23.6 & 62.0 & 16.6 & \best{29.0} & \best{19.9} & 61.0 & 24.7 & \second{25.2} & 22.9 \\
GPT 5.1 & \second{66.6} & 29.2 & \best{39.9} & 29.4 & 64.4 & 28.0 & \best{38.2} & \best{29.5} & 60.6 & 15.5 & \second{28.5} & \second{19.5} & 63.1 & 25.8 & \best{33.3} & \best{26.4} \\
\midrule
\textit{Avg.} & 63.4 & 38.0 & 25.4 & 26.5 & 64.5 & 29.5 & 22.5 & 22.5 & 61.0 & 21.3 & 17.2 & 15.6 & 62.5 & 29.2 & 20.3 & 21.0 \\ 
\textit{Std.} & 2.9 & 7.2 & 7.9 & 5.4 & 2.4 & 9.8 & 7.4 & 6.3 & 1.6 & 4.7 & 6.8 & 3.0 & 1.8 & 4.4 & 5.9 & 3.6 \\
\bottomrule
\end{tabular}
}

\caption{Detailed category-wise evaluation results on \dataset.}
\label{tab:detail_results_by_category}
\end{table*}

\begin{table*}[t]
\centering
\resizebox{0.9\textwidth}{!}{
\begin{tabular}{l|cccc|c}
\toprule
Model & Functionality & Constraint & Interaction & Content & Overall \\
\midrule
MiMo-V2-Flash & 22.2/20.1/19.3 & 50.5/49.7/46.6 & 33.3/31.1/30.3 & 16.2/16.7/15.8 & 44.3/30.4/31.9 \\ 
Step-3.5-Flash & 33.5/25.6/26.4 & 61.4/56.9/55.6 & 39.2/36.9/35.9 & 30.3/33.3/31.2 & 53.4/38.7/40.9 \\ 
GPT-5.1 & 42.8/54.8/42.0 & 57.0/76.5/59.2 & 55.9/65.7/56.0 & 36.5/46.2/38.2 & 47.0/63.4/46.8 \\ 
Claude Sonnet 4.5 & 45.0/32.3/34.5 & 69.1/72.6/66.7 & 44.8/42.8/41.4 & 32.3/32.7/30.7 & 61.0/47.4/49.2 \\ 
\bottomrule
\end{tabular}%
}
\caption{
Detailed results on \dataset in the oracle setting.
}
\label{tab:detail_oracle_results}
\vspace{-0.5em}
\end{table*}

\section{More Related Work}
\label{appsec:more_related_work}

\textbf{Web Application Generation.}
LLMs have catalyzed a paradigm shift in programming, evolving from basic code completion~\citep{fried2023incoder, li2023starcoder} to coding agents~\citep{yang2024sweagent, zhang-etal-2024-codeagent, gao2025trae} and project synthesis from scratch~\citep{zan2024codes, zhao2024commit0}. These advances lay a crucial foundation for the application of LLM in web application generation. 
Early studies focus on the UI-to-Code paradigm~\citep{yun2024web2code, wu-2024-uicoder}, where models replicate web designs from visual screenshots. However, corresponding benchmarks focus on static visual fidelity, such as visual similarity or component matching~\citep{si2025design2code, xiao2025designbench, lin2025webuibench}, neglecting dynamic interactions~\citep{xiao2024interaction2code}. 
In contrast, Language-to-Code is closer to practical development: users specify desired functionality and appearance, and an agent synthesizes interactive applications~\citep{wan2025tddev}. ArtifactsBench~\citep{zhang2025artifactsbench} introduce interaction-oriented evaluation via interaction screenshots, while WebGen-Bench~\citep{lu2025webgenbench} used web agents to test functionality.
However, these evaluations rely on predefined checklists, limiting their ability to handle open-end development scenarios.

\textbf{Automated Web Testing.}
Web application testing is a critical yet time-consuming phase in software development, motivating the pursuit of automated solutions. 
Early automation methods like Selenium use script tools to execute user operations. While simplifying manual tasks, they tightly bind to the website's DOM structure and event triggers, leading to limited scalability and high maintenance costs~\citep{brisset2022erratum}. 
Training-based methods~\citep{li2019humanoid, lan2024dqt} can learn testing strategies, but they suffer from high training costs and limited generalization to emerging web frameworks.
Recent works~\citep{liu2024gptdroid, hu2025kuitest} investigate multimodal LLMs for testing mobile applications. For web applications, \citet{le2025automated} focus on automating site navigation and form filling. UXAgent~\citep{lu2025uxagent} introduces personalized agents for testing web applications. In this paper, we focus on the automated web application testing. Beyond basic functionality, we explore agents' performance on long-horizon tasks and their ability to detect latent bugs.

\clearpage
\onecolumn

\begin{tcolorbox}[
title = {Prompt for Checklist Generation Agent},
breakable,
fontupper=\small,
fonttitle=\small
]
\# Role\\
You are a Senior Software Quality Assurance Engineer who can read a user instruction and immediately produce a complete, executable UI/UX test checklist. Your focus is strictly on 'what' the application should do (the features), not 'how' it should be built (the technical implementation).\\

\# Task\\
Directly generate executable test checklist. Decompose the user instruction into structured, testable items.\\

Each item must be:\\
- **Specific**: Clear action and expected outcome.\\
- **Binary**: PASS or FAIL (no ambiguity).\\
- **Debuggable**: Failure indicates exactly what's missing.\\

Your checklist will be used to:\\
1. Test web applications. Produce PASS/FAIL results for each test item.\\
2. Generate detailed bug reports identifying which requirement failed and why.\\

\#\# Checklist Item Category\\
1. Functionality (FT)\\
\hspace*{1em}* Focus: Core user tasks and workflows that must succeed when inputs are valid.\\
\hspace*{1em}* Scope: What happens when everything goes right?\\
\hspace*{1em}* Example: "User can submit a search query", "User can add an item to the cart".\\
2. Constraint (CS)\\
\hspace*{1em}* Focus: Rules, validations, state invariants, and conflict-prevention logic that prevent the system from entering invalid or contradictory states.\\
\hspace*{1em}* Scope: What prevents the user from doing the wrong thing? What happens with conflicting data?\\
\hspace*{1em}* Examples: "Meeting room cannot be booked if already occupied.", "Cannot submit form with empty required fields."\\
3. Interaction (IX)\\
\hspace*{1em}* Focus: Dynamic behaviors and system responses to user actions (non-functional visual/state changes, user experience).\\
\hspace*{1em}* Scope: How does the interface respond to events like clicks, hover?\\
\hspace*{1em}* Examples: "Show success toast after reservation is created."\\
4. Content (CT)\\
\hspace*{1em}* Focus: The relevance and integrity of text, data, and media (images, icons, videos). Content must strictly align with the instruction's theme/purpose.\\
\hspace*{1em}* Scope: Is the displayed information relevant, and fully functional?\\
\hspace*{1em}* Examples: "All displayed images must be directly relevant to the theme of 'iPhone'."\\

\#\# Default Data\\
Assume the application has default data (e.g., pre-existing products in a store). Do not create new data for testing; use the default data already present in the application.\\

\# Unified Checklist Item Template\\

\`{}\`{}\`{}markdown\\
- [ ] [ID]: [Test description]\\
\hspace*{1em}- Action: [What to do]\\
\hspace*{1em}- Expected: [What should happen]\\
\`{}\`{}\`{}\\

\# Output Format (Markdown)\\

\`{}\`{}\`{}markdown\\
\# Test Checklist\\

\#\# Functionality\\
- [ ] FT-01: [use unified template]\\
- [ ] FT-02: [use unified template]\\

\#\# Constraint\\
- [ ] CS-01: [use unified template]\\

\#\# Interaction\\
- [ ] IX-01: [use unified template]\\

\#\# Content\\
- [ ] CT-01: [use unified template]\\
\`{}\`{}\`{}\\

\# Rules\\
1. Testable: Every item must produce a clear Pass/Fail result.\\
2. Executable: Quality assurance tester should know exactly what to do.\\
3. Specific for action/expected: Include exact element names, button text, expected messages, etc.\\
4. Concise for description: Test description should be 1-2 lines, action/expected should be brief.\\
5. No Implementation: Specify what the app does, not how it's built (no framework details).\\
6. Desktop Only: Ignore responsive design requirements.\\
7. Max 20 items total: Prioritize core requirements. Keep only what is necessary to satisfy the instruction.\\
8. No Redundancy: Avoid duplicating content or behavior that is covered by other categories (e.g., "success messages" should be included only once). Each checklist item MUST be assigned exactly one primary category (FT / CS / IX / CT), even if it has secondary implications. \\

\# Input\\

\#\# User Instruction\\
\textcolor{blue}{\$instruction}\\

\# Output (Markdown)\\
\end{tcolorbox}
\noindent\begin{minipage}{\textwidth}
\captionof{figure}{Prompt for Checklist Generation Agent.}
\label{fig:prompt_checklist_generation}
\end{minipage}

\begin{tcolorbox}[
title = {Prompt for Defect Detection Agent},
breakable,
fontupper=\small,
fonttitle=\small
]
\# Role\\
You are an expert Quality Assurance Test Engineer specializing in automated UI/UX testing. Your task is to validate a web application against a provided checklist. You must systematically execute actions, verify results, and update the checklist status.\\

\# Execution Standards\\

\#\# 1. Interaction Strategy\\
- Tool Use: Use **Playwright tools** to interact with the DOM. Disallow the use of \`{}Bash\`{}, \`{}Read\`{}, and \`{}Write\`{} tools to operate web pages.\\
- DOM-Only: Do NOT use screenshots or visual validation. Rely on DOM attributes (text, id, class, accessibility roles) for verification.\\
- Integrity: Execute all items; never skip. If an item cannot be done, mark FAIL with a concrete reason (no hallucination).\\
- Batching: For pure data entry (e.g., filling a form), you may combine multiple \`{}fill/select\`{} actions into a single code block to save time.\\
- Limited Budget: The entire execution process must operate within a limited budget of turn/tool-call (max 150 times total). Plan first, and execute with as few operations as possible.\\
- Navigation: Only navigate if the checklist item explicitly requires it. Disable page refresh operations unless the page crashes.\\

\#\# 2. Verification Logic\\
- Strict Verification: Compare the \`{}`Actual\`{} behavior of the page against the \`{}Expected\`{} field in the checklist.\\
- Pass: The feature works exactly as described.\\
- Fail: Any deviation (missing element, wrong text, no response, error message) is a FAIL.\\

\#\# 3. Workflow\\
1. Initialize: Navigate to the Target URL.\\
2. Iterate: Go through the Checklist items.\\
3. Execute: Perform the `Action` defined in the item.\\
4. Verify: Check if the `Expected` result is met.\\
5. Record: Update the item's status immediately in your internal memory.\\

\# Output Format (Markdown)\\
You must output the Full Checklist with updated statuses. Do not summarize; return the complete list.\\

\#\# Unified Result Item Template\\

If PASS: Change \`{}- [ ]\`{} to \`{}- [X]\`{} to mark the test as passed.\\

\`{}\`{}\`{}markdown\\
- [X] TEST-ID: [original Description]\\
\hspace*{1em}- Action: [original Action]\\
\hspace*{1em}- Expected: [original Expected]\\
\`{}\`{}\`{}\\

If FAIL: Keep \`{}- [ ]\`{} and append a `Bug Report` block immediately after the test item.\\
  
\`{}\`{}\`{}markdown\\
- [ ] TEST-ID: [original Description]\\
\hspace*{1em}- Action: [original Action]\\
\hspace*{1em}- Expected: [original Expected]\\
\hspace*{1em}- Bug Report:\\
\hspace*{2em}- Issue: [Specific problem type: e.g., Unresponsive Button, Incorrect Form Submission, Element Occlusion]\\
\hspace*{2em}- Actual: [Quote the observed deviation: e.g., Button does not trigger the expected modal, Button text overlaps with icon]\\
\`{}\`{}\`{}\\

\#\# Output Template\\

\`{}\`{}\`{}markdown\\
\# Test Result\\

\#\# Functionality\\
\text{[use unified result item template for each FT-xx]}\\
\text{[use unified result item template for each FT-xx]}\\

\#\# Constraint\\
\text{[use unified result item template for each CS-xx]}\\

\#\# Interaction\\
\text{[use unified result item template for each IX-xx]}\\

\#\# Content\\
\text{[use unified result item template for each CT-xx]}\\
\`{}\`{}\`{}

\# Input\\

\#\# User Instruction\\
\textcolor{blue}{\$instruction}\\

\#\# Application URL\\
\textcolor{blue}{\$server\_url}\\

\#\# Test Checklist\\
\`{}\`{}\`{}markdown\\
\textcolor{blue}{\$checklist}\\
\`{}\`{}\`{}\\

\# Output\\
\end{tcolorbox}
\noindent\begin{minipage}{\textwidth}
\captionof{figure}{Prompt for Defect Detection Agent.}
\label{fig:prompt_defect_detection}
\end{minipage}

\clearpage
\begin{tcolorbox}[
title = {Prompt for Test Item Matching},
breakable,
fontupper=\small,
fonttitle=\small
]
Given a list of predicted checklist items and a list of gold checklist items. You are required to align predicted (model-generated) test items to gold (human-labeled) test items for the same web instruction.\\

Instruction:\\
\`{}\`{}\`{}\\
\textcolor{blue}{\$instruction}\\
\`{}\`{}\`{}\\

Gold Test Items (\`{}"gold\_id": "description"\`{}):\\
\`{}\`{}\`{}\\
\textcolor{blue}{\$gold\_items}\\
\`{}\`{}\`{}\\

Predicted Test Items (\`{}"pred\_id": "description"\`{}):\\
\`{}\`{}\`{}\\
\textcolor{blue}{\$pred\_items}\\
\`{}\`{}\`{}\\

Goal:\\
For each predicted item, decide if it corresponds to exactly one gold item describing the same requirement/behavior. Produce a one-to-one mapping; unmatched predictions should map to None.\\

Matching rules:\\
1. Mapping constraint: each predicted item maps to AT MOST ONE gold item; each gold item MAY be assigned to MULTIPLE predicted items.\\
2. Prioritize intent over wording: if a predicted item is more specific/less specific but clearly covers the same user requirement, match it; otherwise, leave it unmatched.\\
3. Do NOT force matches: if no gold item cleanly aligns, use None.\\
4. Preserve predicted order: output tuples follow the input predicted sequence; length of output list equals number of predicted items.\\

Output Format (Markdown)\\
\text{[}("pred\_id\_1", "gold\_id" or None), ("pred\_id\_2", "gold\_id" or None), ...\text{]}\\

DO NOT PROVIDE ANY OTHER OUTPUT TEXT OR EXPLANATION. Only output the List. Output:\\
\end{tcolorbox}
\noindent\begin{minipage}{\textwidth}
\captionof{figure}{Prompt for Test Item Matching.}
\label{fig:prompt_match_item}
\end{minipage}

\twocolumn

\end{document}